\documentclass[useAMS, usenatbib, fleqn]{mn2e}
\pdfoutput=1
\usepackage{aecompl}
\usepackage{amsmath}
\usepackage{amssymb}
\usepackage{color}
\usepackage{graphicx}
\usepackage{multirow}
\usepackage{times}
\usepackage{url}
\usepackage{hyperref}
\hypersetup{colorlinks=true,citecolor=blue,linkcolor=blue,filecolor=blue,urlcolor=blue}

\newcommand*\diff{\mathop{}\!\mathrm{d}}

\title[Simulating the dust content of galaxies]
    {Simulating the dust content of galaxies: successes and failures}
\author[R.~McKinnon et al.]
    {\parbox{18cm}{Ryan McKinnon$^{1}$\thanks{E-mail: ryanmck@mit.edu},
     Paul Torrey$^{1,2}$,
     Mark Vogelsberger$^{1}$,
     Christopher C.~Hayward$^{2,3}$\thanks{Moore Prize Postdoctoral Scholar in Theoretical Astrophysics}, \\
     and Federico Marinacci$^{1}$}\vspace{0.3cm}\\
     $^{1}$Department of Physics and Kavli Institute for Astrophysics and Space Research,
           Massachusetts Institute of Technology,
           Cambridge, MA 02139, USA \\
     $^{2}$TAPIR, California Institute of Technology, Pasadena, CA 91125, USA \\
     $^{3}$Harvard-Smithsonian Center for Astrophysics, 60 Garden St., Cambridge, MA 02138, USA
    }
\begin{document}

\date{Accepted ???. Received ???; in original form ???}

\pagerange{\pageref{firstpage}--\pageref{lastpage}}
\pubyear{???}

\maketitle

\label{firstpage}

\begin{abstract}
We present full volume cosmological simulations using the moving-mesh
code \textsc{arepo} to study the coevolution of dust and galaxies.  We extend
the dust model in \textsc{arepo} to include thermal sputtering of grains and
investigate the evolution of the dust mass function, the cosmic distribution of
dust beyond the interstellar medium, and the dependence of dust-to-stellar mass
ratio on galactic properties.  The simulated dust mass function is
well-described by a Schechter fit and lies closest to observations at $z = 0$.
The radial scaling of projected dust surface density out to distances of $10 \,
\text{Mpc}$ around galaxies with magnitudes $17 < i < 21$ is similar to that
seen in Sloan Digital Sky Survey data, albeit with a lower
normalisation.  At $z = 0$, the predicted dust density of
$\Omega_\text{dust} \approx 1.3 \times 10^{-6}$ lies in the range of
$\Omega_\text{dust}$ values seen in low-redshift observations.  We find that
dust-to-stellar mass ratio anti-correlates with stellar mass for galaxies
living along the star formation main sequence.  Moreover, we estimate the
$850 \, \mu\text{m}$ number density functions for simulated galaxies
and analyse the relation between dust-to-stellar flux and mass ratios at $z =
0$.  At high redshift, our model fails to produce enough dust-rich galaxies,
and this tension is not alleviated by adopting a top-heavy initial mass
function.  We do not capture a decline in $\Omega_\text{dust}$ from $z = 2$ to
$z = 0$, which suggests that dust production mechanisms more strongly dependent
on star formation may help to produce the observed number of dusty galaxies
near the peak of cosmic star formation.
\end{abstract}


\section{Introduction}

The dust content of high-redshift galaxies provides insight into star formation
and metal enrichment at early times, and the abundance of dusty, starburst
galaxies at submillimetre wavelengths \citep{Smail1997, Barger1998, Hughes1998,
Blain1999c, Eales1999, Scott2002} has implications for theories of galaxy
formation and evolution \citep{Blain1999b, Chary2001, Dunne2003a, Hayward2013,
Casey2014}.  Models are challenged to explain the presence of such galaxies and
the key environmental factors that contribute to their growth.  Highlighting
this difficulty, there are recent observations of dusty galaxies at extremely
high redshift, including HFLS3 at $z = 6.34$ with dust mass $M_\text{dust} =
1.3 \times 10^{9} \, \text{M}_\odot$ \citep{Riechers2013}, A1689-zD1 at $z =
7.5$ with $M_\text{dust} = 4 \times 10^{7} \, \text{M}_\odot$
\citep{Watson2015}, and two gravitationally-lensed dusty sources at $z = 5.7$
\citep{Vieira2013, Hezaveh2013, Weiss2013}.

While (ultra)luminous infrared galaxies are roughly a thousand times more
abundant at high redshift ($z \sim 2-3$) than at low redshift
\citep{Chapman2005, Lagache2005}, not all high-redshift star-forming galaxies
are dust-rich.  A prominent example is Himiko, a $z = 6.595$ galaxy with
star-formation rate~(SFR) roughly $100 \, \text{M}_\odot \, \text{yr}^{-1}$ but
very weak dust emission \citep{Ouchi2013}.  The fact that some actively
star-forming galaxies are dust-rich while others are dust poor motivates a
closer study of high-redshift galaxies to better understand their formation.

One important statistic is the dust mass function~(DMF), whose evolution in
time tracks dust growth across large populations of galaxies.  The DMF was
first measured at low redshift as part of the SCUBA Local Universe Galaxy
Survey \citep{Dunne2000, Dunne2001, Vlahakis2005}.  Evolution in the DMF has
been studied over $0 < z < 1$ \citep{Eales2009, Dunne2011, Clemens2013}, and
observations from the Herschel ATLAS \citep{Eales2010} find that the largest
galaxies at $z = 0.5$ contained roughly five times more dust than those in the
local universe \citep{Dunne2011}.  The DMF has been estimated for $1 < z < 5$
using observations and number counts of submillimetre galaxies, with dust-rich
galaxies showing the most change compared to the present day
\citep{Dunne2003a}.  Given the correlation between bolometric luminosity or SFR
and dust obscuration \citep{Wang1996, Adelberger2000, Reddy2006}, the evolution
of the DMF is connected to changes in the luminosity function.  Luminosities of
star-forming, dust-obscured galaxies at high redshift have been analysed in
survey data \citep{Reddy2006, Dey2008, Magdis2012, Magnelli2012, LoFaro2013,
Sklias2014}, and galaxies at $z \sim 2$ are noticeably more luminous than their
local counterparts for fixed dust obscuration \citep{Reddy2010}.  However, the
DMF remains less studied than statistics like the galaxy stellar mass function,
particularly in the high-redshift regime where observations are challenging.

To approach this problem from a theoretical perspective, a number of models
have been developed to study the population of submillimetre galaxies.  Many of
these models employ radiative transfer to self-consistently track absorption
and reradiation of stellar light by dust (e.g.~using the \textsc{grasil}
\citep{Silva1998}, \textsc{sunrise} \citep{Jonsson2006}, \textsc{radishe}
\citep{Chakrabarti2009}, or \textsc{art$^2$} \citep{Yajima2012} codes) and to
estimate submillimetre flux densities and number counts.  Radiative transfer
can be combined with semi-analytic models \citep{Baugh2005, Swinbank2008} or
hydrodynamical simulations of galaxies \citep{Chakrabarti2008, Narayanan2009,
Narayanan2010, Hayward2011, Hayward2012, Hayward2013} to investigate how
various galactic properties impact submillimetre galaxies.

Such simulations have shown that flux densities in the SCUBA $850 \,
\mu\text{m}$ \citep{Holland1999} and AzTEC $1.1 \, \text{mm}$
\citep{Wilson2008} bands can be well estimated from a galaxy's SFR and dust
mass \citep{Hayward2011}.  This agrees with findings that dust obscuration
correlates with SFR \citep{Adelberger2000}.  Furthermore, a top-heavy initial
mass function~(IMF), at least in starbursts, may help to explain number counts
of submillimetre galaxies and their dust content~\citep{Baugh2005,
Swinbank2008, Michalowski2010b}.

The predictive capability of such semi-analytic and radiative transfer models
motivates the inclusion of dust physics directly into galaxy formation
simulations where more diverse samples of galaxies can be studied and the
evolution of quantities like the DMF can be traced.  The direct treatment of
dust in cosmological simulations of uniform volumes provides the opportunity to
investigate which environmental factors most contribute to the formation of
dusty, submillimetre galaxies.  It also enables comparison with a variety of
observations that cannot be fully tested in simulations of individual galaxies.
These include the radial scaling of projected dust surface density around
galaxies to distances of several Mpc \citep{Menard2010}, the relation between
SFR, stellar mass, and dust mass at low redshift and out to $z = 2.5$
\citep{DaCunha2010, Dunne2011, Skibba2011, Bourne2012, Cortese2012, Davies2012,
Rowlands2012, Smith2012, Clemens2013, Santini2014, RemyRuyer2015}, and
estimates of the cosmic dust density parameter $\Omega_\text{dust}$ and its
evolution \citep{Fukugita2004, Driver2007, Menard2010, Dunne2011, Fukugita2011,
DeBernardis2012, Menard2012, Clemens2013, Thacker2013}.

In previous work \citep[][hereafter M16]{McKinnon2016}, we introduced a dust
model accounting for the production of dust through stellar evolution,
accretion in the interstellar medium~(ISM) via collisions with gas-phase
metals, and non-thermal sputtering in supernova~(SN) shocks that returned dust
to the gas phase, and performed zoom-in simulations of a suite of eight Milky
Way-sized haloes.  Here, we extend the dust model from M16 and perform the
first cosmological simulations of galaxy populations in which dust is directly
treated.

This paper is organized as follows.  In Section~\ref{SEC:methods}, we describe
the inclusion of new physics into our existing galaxy formation model and
detail the initial conditions used for our simulations.  In
Section~\ref{SEC:results}, we present our results and compare with existing
data, and, in Section~\ref{SEC:discussion}, we discuss the implications of our
findings in a broader context.  Finally, Section~\ref{SEC:conclusions}
summarizes our results and offers an outlook on future work.

\section{Methods}\label{SEC:methods}

We perform cosmological simulations using the moving-mesh code \textsc{arepo}
\citep{Springel2010}.  The simulations incorporate the galaxy formation physics
described in \citet{Vogelsberger2013}.  Briefly, this galaxy formation model
includes gravity, hydrodynamics, primordial and metal-line cooling
\citep{Wiersma2009a}, black hole growth \citep{Sijacki2007}, star formation
\citep{Springel2003}, stellar evolution, chemical enrichment tracking nine
elements (H, He, C, N, O, Ne, Mg, Si, and Fe), and stellar and active galactic
nuclei feedback.  It has been used in previous cosmological simulations,
including the Illustris simulation, that trace the evolution of the galaxy
stellar mass function, luminosity function, mass-metallicity relation, and
other quantities and shows broad agreement with observations
\citep{Vogelsberger2012, Torrey2014, Vogelsberger2014a, Vogelsberger2014b,
Genel2014}.  In addition to this galaxy formation model, we employ a modified
version of the dust model from M16, which is described below and changes our
treatment of dust in the circumgalactic medium~(CGM).

\subsection{Dust Model}

The dust model in M16 accounts for dust production from aging stellar
populations, grain growth, destruction in SN shocks, and the advection and
transport of dust in galactic winds.  Dust is injected into the ISM as stars
evolve off the main sequence, with dust masses calculated using stellar
nucleosynthetic yields and estimated grain condensation efficiencies.
The time-scale for grain growth through collisions between gas-phase atoms and
grains depends on local gas density and temperature, while the time-scale for
dust destruction through SN sputtering scales inversely with the local SNe
rate.  Here, we also model the evolution of dust in galactic haloes.  The
physics of dust grains in hot gas has been studied in detail and includes
sputtering, cooling, and grain-grain collisions \citep{Ostriker1973, Burke1974,
Salpeter1977, Barlow1978, Draine1979b, Itoh1989, Tielens1994, Dwek1996,
Smith1996}.  Thermal sputtering allows for the erosion of dust grains by
energetic atoms, and it can limit the depletion of gas-phase metals onto grains
in hot parts of a galactic halo \citep{Burke1974, Barlow1978, Draine1979b} and
possibly enrich the intergalactic medium with metals \citep{Bianchi2005}.
Thermal sputtering affects grain lifetimes and gas cooling in the intracluster
medium \citep{Yahil1973, Dwek1992, McGee2010}.  Hydrogen and helium are the
main sputtering agents, and predictions of thermal sputtering rates indicate
that sputtering overwhelms dust growth via accretion of gas-phase atoms for
$10^5 \, \text{K} < T < 10^9 \, \text{K}$ \citep{Draine1979b}.  The strength of
thermal sputtering is expected to decline sharply below $T \sim 10^{6} \,
\text{K}$ \citep{Ostriker1973, Barlow1978, Draine1979b, Tielens1994,
Nozawa2006}.

We outline below the inclusion of thermal sputtering into the dust model used
in M16.  It is expected that other grain destruction mechanisms, like
grain-grain collisions and cosmic ray-driven sputtering, are subdominant
compared to non-thermal SN shocks and thermal sputtering \citep{Barlow1978,
Draine1979a, Draine1979b, Jones1994}.  We follow thermal sputtering
prescriptions as used in previous galaxy modelling \citep{Tsai1995,
Hirashita2015} for simplicity of implementation.

Following Equation~14 in \citet{Tsai1995}, we estimate the sputtering
rate for a grain of radius $a$ in gas of density $\rho$ and temperature $T$ as
\begin{equation}
\frac{\diff a}{\diff t} = -(3.2 \times 10^{-18} \, \text{cm}^{4} \, \text{s}^{-1}) \left( \frac{\rho}{m_\text{p}} \right) \left[ \left( \frac{T_0}{T} \right)^\omega + 1 \right]^{-1},
\label{EQN:adot}
\end{equation}
where $m_\text{p}$ is the proton mass, $\omega = 2.5$
controls the low-temperature scaling of the sputtering rate, and $T_0 = 2
\times 10^{6} \, \text{K}$ is the temperature above which the sputtering rate
is approximately constant.  This empirical fitting formula approximately
captures the temperature dependence of sputtering rates derived in theoretical
calculations of collisions between spherical grains and impinging gas
particles, which we outline in Appendix~\ref{SEC:appendix_sputtering}
\citep{Barlow1978, Draine1979b, Tielens1994}.  The associated sputtering
time-scale for the grain is given by Equation~15 in \citet{Tsai1995},
\begin{equation}
\tau_\text{sp} = a \left| \frac{\diff a}{\diff t} \right|^{-1} \approx (0.17 \, \text{Gyr}) \left( \frac{a_{-1}}{\rho_{-27}} \right) \left[ \left( \frac{T_0}{T} \right)^\omega + 1 \right],
\label{EQN:tau_sp}
\end{equation}
where $a_{-1}$ is the grain size in units of $0.1 \mu\text{m}$ and $\rho_{-27}$
is the gas density in units of $10^{-27} \, \text{g} \, \text{cm}^{-3}$, which
corresponds to an effective number density of $n \approx 6 \times 10^{-4} \,
\text{cm}^{-3}$.  This time-scale is similar to the approximate sputtering
time-scale given in Equation~44 of \citet{Draine1979b}, where detailed
projectile calculations were performed.  Given a grain of constant internal
density $\rho_\text{g}$ and mass $m_\text{g} = 4 \pi a^3 \rho_\text{g} / 3$,
Equation~(\ref{EQN:tau_sp}) implies that mass changes according to the
time-scale $|m/\dot{m}| = \tau_\text{sp}/3$.  In our model, we track the total
dust mass for five chemical species (C, O, Mg, Si, and Fe) within each gas
cell, but we do not track the grain size distribution.  To account for the
effect of thermal sputtering on $M_\text{$i$,dust}$, the species $i$ dust mass
within each gas cell, during every time-step we calculate the dust mass loss
rate
\begin{equation}
\left( \frac{\diff M_\text{$i$,dust}}{\diff t} \right)_\text{sp} = - \frac{M_\text{$i$,dust}}{\tau_\text{sp}/3},
\end{equation}
where $\tau_\text{sp}$ is computed using Equation~(\ref{EQN:tau_sp}) and the
local gas density and temperature.  We fix the grain radius $a = 0.1 \,
\mu\text{m}$ as our model is not equipped to sample across a grain size
distribution.  This choice for $a$ is motivated by the facts that the grain
size distribution for dust produced by AGB stars is thought to peak near $0.1
\, \mu\text{m}$ \citep{Groenewegen1997, Winters1997, Yasuda2012, Asano2013b}
and SNe are expected to form grains with $a \gtrsim 0.01 \, \mu\text{m}$
\citep{Bianchi2007, Nozawa2007}.  We show in Appendix~\ref{SEC:appendix_grain}
that our results do not strongly depend on this grain size assumption.

Combining with the dust accretion and SN-based destruction rates from
Equations~4 and~5 of M16, the net rate of dust mass change is given by
\begin{equation}
\frac{\diff M_\text{$i$,dust}}{\diff t} = \left( 1 - \frac{M_\text{$i$,dust}}{M_\text{$i$,metal}} \right) \left( \frac{M_\text{$i$,dust}}{\tau_\text{g}} \right) - \frac{M_\text{$i$,dust}}{\tau_\text{d}} - \frac{M_\text{$i$,dust}}{\tau_\text{sp}/3},
\end{equation}
where $M_\text{$i$,metal}$ is the metal mass of species $i$ in the cell and the
growth and destruction time-scales $\tau_\text{g}$ and $\tau_\text{d}$ depend on
the local density, temperature, and Type II SN rate, as indicated by
Equations~5 and~7 in M16.  This dust mass rate is computed on a
cell-by-cell basis for every species and used to update dust masses in every
time-step.

We summarize the set of parameters and quantities that characterize our
fiducial dust model in Table~\ref{TAB:parameters}.  The dust model used in this
work differs from that of M16 in two respects: (i) it includes thermal
sputtering, and (ii) the dust growth parameters have been changed slightly to
follow from \citet{Hirashita2000}, which offers a more detailed analysis of
dust growth time-scales in molecular clouds.  As shown in
Section~\ref{SEC:results}, this latter change was adopted to lessen depletion
at low redshift compared to the M16 model and better match the observed DMF and
cosmic dust density parameter.

\begin{table*}
\centering
\caption{Summary of parameters used in various components of the full fiducial
dust model.  The dust condensation efficiencies that we use to compute dust
produced from stellar evolution are unchanged from those given in Table~2 of
M16.  The dust accretion parameters, which affect the growth time-scale
calculated in Equation~5 of M16, differ slightly from those used in M16 and are
based on Equation~12 in \citet{Hirashita2000}.}
\begin{tabular}{lll}
\hline
Parameter & Value & Description \\
\hline
\hline
thermal sputtering & & \\
\hline
$a$ & $0.1$ & grain radius, in units of $[\mu\text{m}]$ \\
\hline
dust accretion & & \\
\hline
$\rho^\text{ref}$ & $2.3 \times 10^{-22}$ & reference density roughly corresponding to $n_\text{H} = 100 \, \text{cm}^{-3}$, in units of $[\text{g} \, \text{cm}^{-3}]$\\
$T^\text{ref}$ & $50$ & reference temperature, in units of $[\text{K}]$ \\
$\tau_\text{g}^\text{ref}$ & $0.4$ & dust growth time-scale when $T = T^\text{ref}$ and $\rho = \rho^\text{ref}$, in units of $[\text{Gyr}]$ \\
\hline
SN-based destruction & & \\
\hline
$E_\text{SNII,$51$}$ & 1.09 & energy per SN II, in units of $[10^{51} \, \text{erg}]$ \\
\hline
\end{tabular}
\label{TAB:parameters}
\end{table*}

\begin{table*}
\centering
\caption{Summary of simulation parameters and resolutions used in this work.
Here, $N$ is the total particle number, including equal numbers of dark matter
and gas cells to start; $\epsilon$ is the maximum physical gravitational
softening length, attained at $z = 1$; $m_\text{dm}$ is the dark matter
resolution; and $m_\text{gas}$ is the target gas mass for each cell in the
(de-)refinement scheme.  The last column describes the physics for each
simulation.  The M16 model refers to dust model used in M16, which lacked
thermal sputtering and adopted dust growth parameters $\tau^\text{ref} = 0.2 \,
\text{Gyr}$, $\rho^\text{ref} = 2.3 \times 10^{-24} \, \text{g} \,
\text{cm}^{-3}$, and $T^\text{ref} = 20 \, \text{K}$, leading to stronger ISM
dust growth.  The top-heavy IMF run uses a pure power-law IMF of the form
$\Phi(m) \propto m^{-1.3}$ over the mass range from $0.1 \, \text{M}_\odot$ to
$100 \, \text{M}_\odot$.  It adopts fiducial dust physics.}
\begin{tabular}{lllllll}
\hline
Name & Volume & $N$ & $\epsilon$ & $m_\text{dm}$ & $m_\text{gas}$ & physics \\
& $[(h^{-1} \, \text{Mpc})^3]$ & & $[h^{-1} \, \text{kpc}]$ & $[h^{-1} \, \text{M}_\odot]$ & $[h^{-1} \, \text{M}_\odot]$ \\
\hline
\hline
L25n128 & $25^3$ & $2 \times 128^3$ & $2.5$ & $5.26 \times 10^{8}$ & $9.82 \times 10^{7}$ & fiducial \\
L25n256 & $25^3$ & $2 \times 256^3$ & $1.25$ & $6.58 \times 10^{7}$ & $1.23 \times 10^{7}$ & fiducial \\
L25n512 & $25^3$ & $2 \times 512^3$ & $0.625$ & $8.22 \times 10^{6}$ & $1.53 \times 10^{6}$ & fiducial \\
\hline
M16 model & $25^3$ & $2 \times 256^3$ & $1.25$ & $6.58 \times 10^{7}$ & $1.23 \times 10^{7}$ & model used in M16 (see caption) \\
top-heavy IMF & $25^3$ & $2 \times 256^3$ & $1.25$ & $6.58 \times 10^{7}$ & $1.23 \times 10^{7}$ & IMF has the form $\Phi(m) \propto m^{-1.3}$ \\
\hline
\end{tabular}
\label{TAB:simulations}
\end{table*}

\subsection{Initial Conditions and Simulations}\label{SEC:ICs}

We simulate uniformly-sampled cosmological volumes of comoving side length $L =
25 \, h^{-1} \, \text{Mpc}$ with combined gas and dark matter particle numbers
of $2 \times 128^3$, $2 \times 256^3$, and $2 \times 512^3$.  The gravitational
softening length is held constant in comoving units until $z = 1$, after which
point it is fixed to the same physical value.  The maximum physical
gravitational softening length is $625 \, h^{-1} \, \text{pc}$ for the run
with $2 \times 512^3$ particles.

We use $\Lambda$CDM cosmological parameters from the reanalysis of Planck data
\citep{PlanckCollaboration2014} by \citet{Spergel2015} of $\Omega_\text{m} =
0.302$, $\Omega_\text{b} = 0.04751$, $\Omega_\Lambda = 0.698$, $\sigma_8 =
0.817$, $n_s = 0.9671$, and $H_0 = 100 \, h \, \text{km} \, \text{s}^{-1} \,
\text{Mpc}^{-1} = 68 \, \text{km} \, \text{s}^{-1} \, \text{Mpc}^{-1}$.
Initial conditions with these parameters are generated at $z = 127$ using
\textsc{music} \citep{Hahn2011} and iterated forward using \textsc{arepo}.  We
use the \textsc{subfind} algorithm \citep{Springel2001, Dolag2009} for
identifying gravitationally-bound structure and calculating gas, stellar, and
dust mass components within galaxies.  Galactic quantities are computed within
twice the stellar half-mass radius.

Table~\ref{TAB:simulations} provides details about the simulations,
including softening lengths and particle resolutions.
Our fiducial simulations are performed at three resolution levels and use the
fiducial galaxy formation and feedback parameters from
\citet{Vogelsberger2013}, including the fiducial \citet{Chabrier2003} IMF.  We
perform two additional simulations to explore how sensitive our results are to
the dust model we use and to the choice of IMF.

First, the ``M16 model'' run uses the dust model from M16, which lacked thermal
sputtering and had dust growth parameters tuned to Milky Way-sized galaxies and
resulting in fairly high depletion.  The dust growth parameters used in this
work and outlined in Table~\ref{TAB:parameters} lead to weaker dust growth than
in M16.

Second, the ``top-heavy IMF'' run uses fiducial dust physics but an IMF of the
form $\Phi(m) \propto m^{-1.3}$, with the same lower mass limit of $0.1 \,
\text{M}_\odot$ and upper mass limit of $100 \, \text{M}_\odot$ as the fiducial
IMF.  The $m > 1 \, \text{M}_\odot$ portion of the \citet{Chabrier2003} IMF
adopts the power law $\Phi(m) \propto m^{-2.3}$, and so the top-heavy IMF we
experiment with increases the exponent by one and extends the power law to the
full mass range.  While our top-heavy IMF is independent of galaxy properties,
we note that previous works have used even more top-heavy IMFs in starbursts,
like $\Phi(m) \propto m^{-1}$ \citep{Baugh2005, Swinbank2008}.  It is thought
that a top-heavy IMF may help form large amounts of dust at high redshift.
Because the stellar feedback prescription in \citet{Vogelsberger2013} is tuned
to a \citet{Chabrier2003} IMF, for our top-heavy simulation we fix the
IMF-dependent quantities in the model of \citet{Springel2003} to their fiducial
values.  To be precise, the parameters $\beta$, the mass fraction of stars with
$m > 8 \, \text{M}_\odot$, and $\epsilon_\text{SN}$, the IMF-averaged energy
returned by supernovae per solar mass of stars formed, are kept at their values
for a \citet{Chabrier2003} IMF to ensure the stellar feedback model is not
strongly affected by the top-heavy IMF.  In theory, to keep the
\citet{Springel2003} model consistent with the top-heavy IMF, we would need to
increase $\beta$, which would in turn affect the SN II rate and SN-driven
sputtering of dust.  However, such changes would affect stellar feedback and
its ability to reproduce the galaxy stellar mass function, which would
complicate the interpretation of our results.  To summarize, the top-heavy run
adopts fiducial dust physics (including thermal sputtering and the
\citet{Hirashita2000} growth time-scale parameterisation) and for mass return
uses a power-law IMF of the form $\Phi(m) \propto m^{-1.3}$, but it keeps the
stellar feedback routines calibrated to the fiducial IMF.

\section{Results}\label{SEC:results}

\begin{figure*}
\centering
\includegraphics{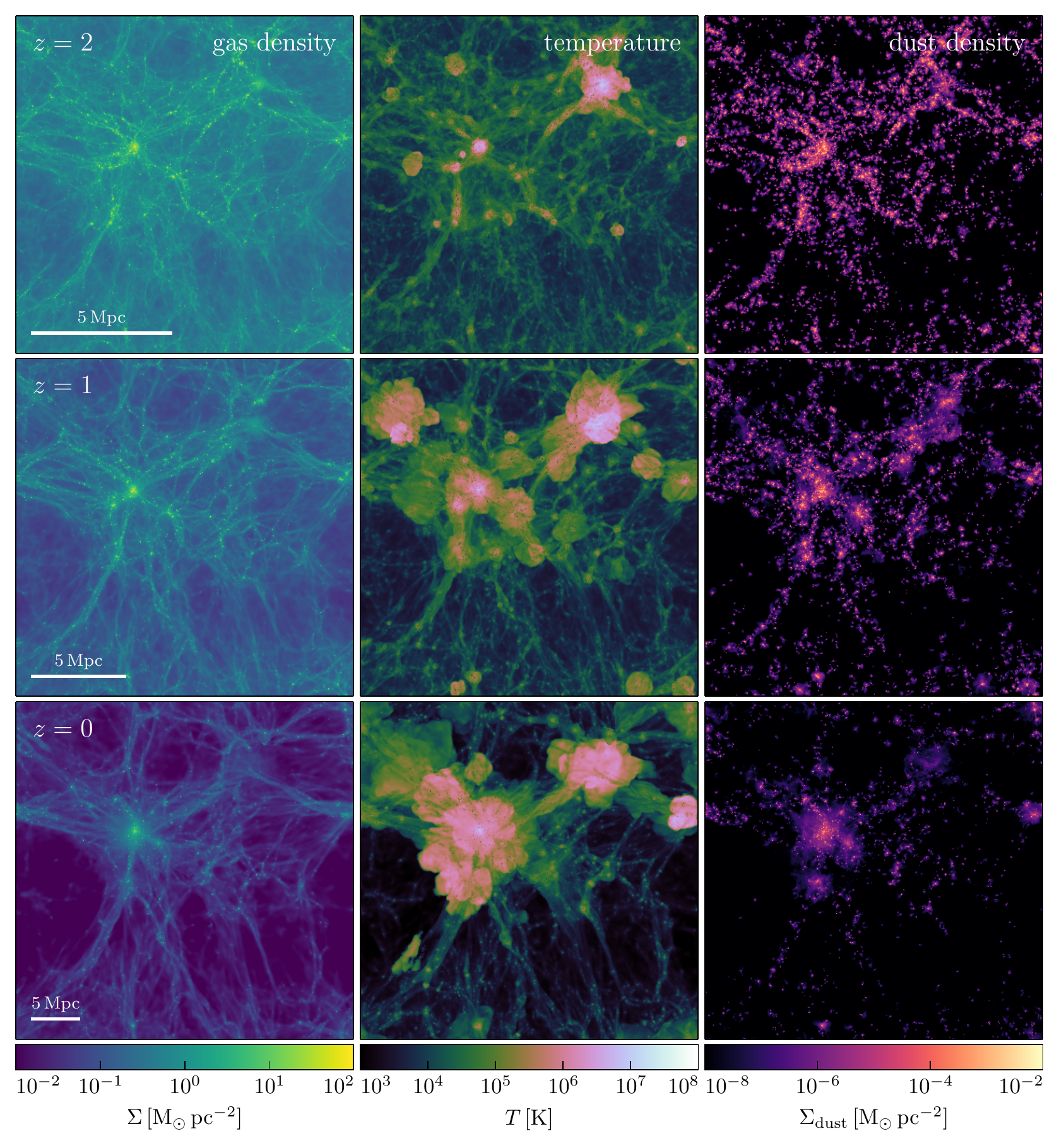}
\caption{Projections of gas density, temperature, and dust density (left,
middle, and right columns) at $z = 2$, $1$, and $0$ (top, middle, and bottom
rows) for the highest resolution simulation.  Densities are given in physical
units, and the scale bar for each redshift indicates a physical distance of $5
\, \text{Mpc}$.  Projections were performed about the centre of the simulated
volume, with a height and width of $25 \, h^{-1} \, \text{Mpc}$ and a depth of
$12.5 \, h^{-1} \, \text{Mpc}$ in comoving units.  The distribution of dust
largely traces that of gas.}
\label{FIG:box_evolution}
\end{figure*}

We first use our highest resolution fiducial simulation to visualize the
distribution of dust and its redshift evolution.
Figure~\ref{FIG:box_evolution} shows projections of gas density, gas
temperature, and dust density through a slice of the full simulation volume at
$z = 2$, $1$, and $0$.  The dust surface density peaks in gas-rich halo
centres, where the efficient production of dust by stars and short time-scales
for grain-atom collisions overcome the presence of SN sputtering.  Comparing
different redshifts shows that the distribution of dust is rearranged through
mergers, as demonstrated by the largest halo at $z = 0$.  It is also clear that
large filaments of cold, diffuse gas far from potential minima -- and thus
sources of dust formation -- have essentially no dust.

Recalling the temperature dependence of the sputtering time-scale given in
Equation~(\ref{EQN:tau_sp}), we can see in Figure~\ref{FIG:box_evolution} that
several of the largest haloes at $z = 0$ have temperatures above $T \approx
10^{6} \, \text{K}$.  At these temperatures, the thermal velocity is
high enough to erode grains.  Lower mass haloes witness lower temperatures
where the thermal sputtering rate falls off sharply.  Regardless of halo size
or temperature, dust in the cool ISM is largely unaffected by thermal
sputtering, and it is interesting to consider the DMF corresponding to this
diverse sample of simulated galaxies.

\subsection{Dust Mass Function}

\begin{figure*}
\centering
\includegraphics{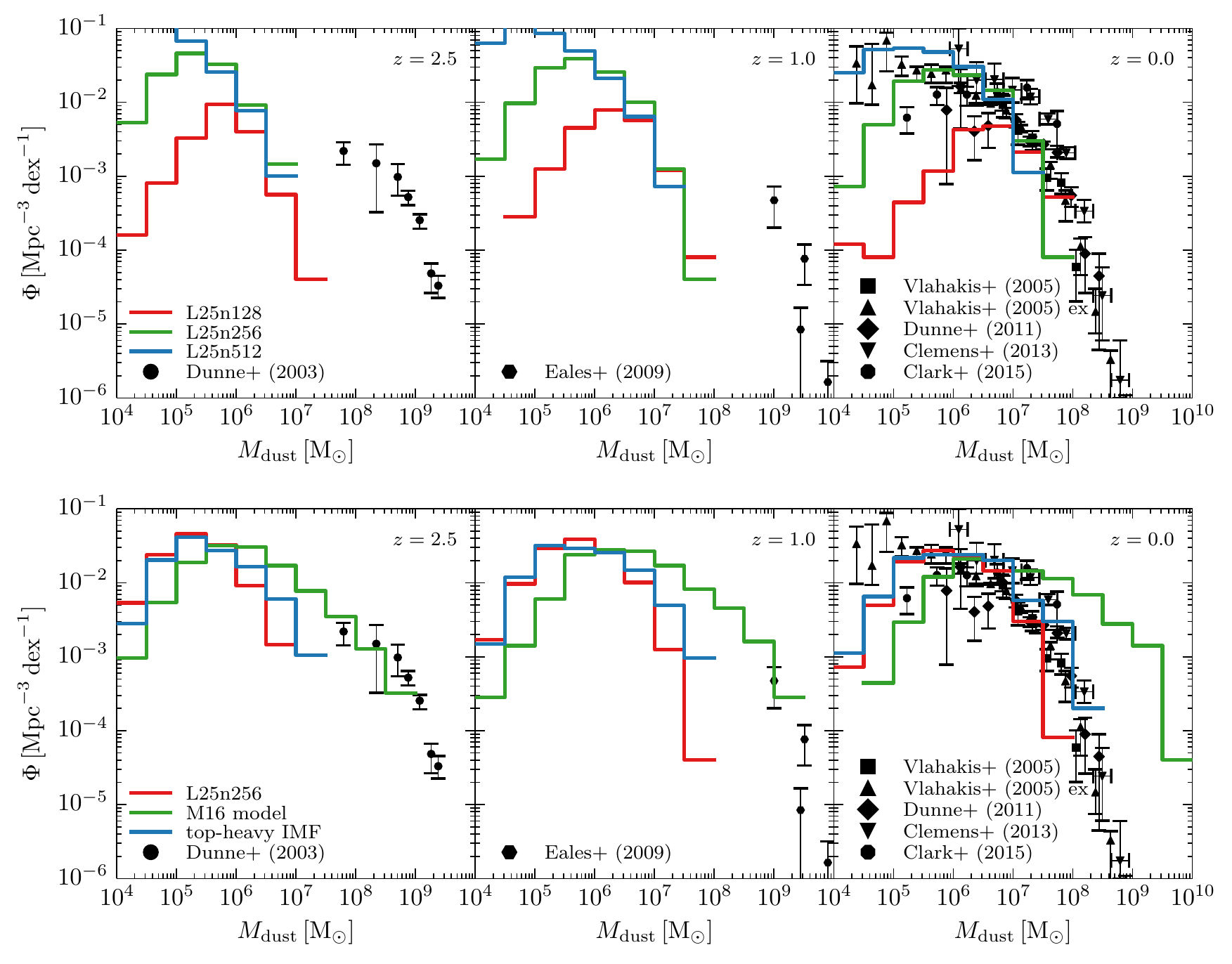}
\caption{Simulated DMFs~(coloured lines) for three resolution levels~(top row)
and model variations~(bottom row) as compared with observations~(black points)
for $z = 2.5$, $1.0$, and $0.0$~(left, middle, and right panels, respectively).
For \citet{Eales2009}, we plot data from $0.6 < z < 1.0$.  From
\citet{Vlahakis2005} we include both the directly measured DMF and the IRAS
PSCz-extrapolated DMF, the latter of which has the suffix ``ex'' in the legend.
For \citet{Dunne2011}, we use the $0.0 < z < 0.1$ set of data and cap the
uncertainty at $1 \, \text{dex}$ for two data points to improve readability.
Observations have been corrected to conform to the cosmology detailed in
Section~\ref{SEC:ICs}.  While the fiducial simulated DMFs offer a
decent fit to observations at $z = 0.0$, they fail to produce an abundance of
dust-rich galaxies at higher redshift.}
\label{FIG:dmf}
\end{figure*}

We plot simulated DMFs at $z = 2.5$, $1.0$, and $0.0$ in Figure~\ref{FIG:dmf}
for our fiducial runs at three different resolutions.  Figure~\ref{FIG:dmf}
also shows the DMFs for our two model variations, one using the dust model from
M16 and the other using a top-heavy IMF.  Dust masses are computed within twice
the stellar half-mass radius.  We compare with a variety of observational data
\citep{Dunne2003a, Vlahakis2005, Eales2009, Dunne2011, Clemens2013, Clark2015},
although we note that high-redshift observations are limited to the massive end
of the DMF.  These data have been corrected to the cosmology described in
Section~\ref{SEC:ICs}.  We also standardise the data to the dust mass
absorption coefficient $\kappa(850 \, \mu\text{m}) = 0.77 \, \text{cm}^{2} \,
\text{g}^{-1}$ adopted in \citet{Dunne2011}.\footnote{This choice of dust mass
absorption coefficient means that dust mass data reported in
\citet{Clemens2013}, which used a value of $\kappa(850 \, \mu\text{m})$ smaller
by roughly a factor of two, have been halved for this work.}  We compare data
from \citet{Dunne2003a} with simulated galaxies at $z = 2.5$, which is the
value used in that work to compute dust masses for galaxies without
spectroscopic redshifts.  However, as noted in Section~3 of \citet{Dunne2003a},
the estimated dust masses are largely insensitive to this choice of redshift.
For the $z = 1.0$ panel, we plot data from \citet{Eales2009} over the range
$0.6 < z < 1.0$.  From \citet{Vlahakis2005}, we include both the
directly-measured DMF and the DMF extrapolated over a larger dust mass range
using IRAS PSCz data, the latter of which is given the suffix ``ex'' in the
legend for Figure~\ref{FIG:dmf}.

While our fiducial model offers a reasonable fit to observed data at $z = 0.0$
down to the resolution limit, it does not reproduce the abundance of high dust
mass galaxies near $z = 2.5$ and $z = 1.0$.  At $z = 2.5$, the number
density of simulated galaxies with $M_\text{dust} \approx 10^{7} \,
\text{M}_\odot$ is similar to that for observed galaxies with $M_\text{dust}
\approx 10^{9} \, \text{M}_\odot$.  Although our simulated value of
$\Phi(M_\text{dust} \approx 2 \times 10^{7} \, \text{M}_\odot)$ increases by
over $1 \, \text{dex}$ from $z = 2.5$ to $1.0$, we still have difficulty
producing enough dust-rich galaxies at $z = 1.0$.  The nature of dust processes
makes the DMF behave in a much more dynamic way than the galaxy stellar mass
function, since there is a diversity of ways for dust to grow (e.g.~stellar
injection of grains and collisions with gas in the ISM) and be destroyed
(e.g.~SN shocks and thermal sputtering).  This same core galaxy formation model
without dust tracking had success in matching the galaxy stellar mass
function's gradual flattening at the low mass end from high to low redshift as
more galaxies gain stellar mass \citep{Vogelsberger2013, Torrey2014,
Vogelsberger2014a, Vogelsberger2014b, Genel2014}.  However, the DMF does not
evolve in such a monotonic fashion: galaxies at $z = 2.5$ tend to be more
dust-rich than at $z = 0.0$ \citep{Dunne2003a}, and, even from $z = 0.5$ to $z
= 0.0$, galactic dust masses decline by about a factor of five
\citep{Dunne2011}.  It is worth noting that for mass bins of width $0.5 \,
\text{dex}$, a DMF value of $\Phi = 10^{-4} \, \text{Mpc}^{-3} \,
\text{dex}^{-1}$ corresponds to roughly two galaxies in our fiducial volume.
Thus, the massive ends of our DMFs are sensitive to Poissonian statistics.  In
Appendix~\ref{SEC:appendix_dmf}, we simulate a volume eight times as large down
to $z = 2.5$ and investigate its DMF.  The greater sample size provided by a
larger volume does not alleviate the absence of very dusty galaxies.
Furthermore, the fiducial runs do not display robust convergence as the
resolution is increased.  At $z = 0.0$, the DMF falls off at the high-mass end
more quickly with increasing resolution.  As a result, number densities
associated with the L25n512 run lie above those for the L25n128 run in some
mass bins, while the trend is reversed in other bins.  We note in
Section~\ref{SEC:rho_dust} that volume-averaged quantities like comoving dust
density display better convergence properties.

In Figure~\ref{FIG:dmf}, the dust model used in M16, which had a stronger dust
growth mechanism and lacked thermal sputtering, differs the most from the
fiducial model at $z = 0.0$ and overproduces dust-rich galaxies.  This is
consistent with the finding in M16 that strong dust growth can overdeplete
gas-phase metals at late times.  These results are largely unaffected by the
inclusion of thermal sputtering since the DMFs in Figure~\ref{FIG:dmf}
isolate dust in the fairly cool ISM.  However, the M16 model does
predict more dust-rich galaxies at $z = 2.5$ than the fiducial model and lies
close to the \citet{Dunne2003a} data.  A dust growth mechanism that allows for
more variation among galaxies of different masses and SFRs may be needed to
form dust-rich galaxies at high redshift but also avoid overproducing dust at
low redshift.

Similarly, the run with a top-heavy IMF, $\Phi(m) \propto m^{-1.3}$, produces
more dust than the fiducial L25n256 run at all redshifts.  This is consistent
with a top-heavy IMF shifting the galaxy stellar mass function towards lower
masses due to shorter average stellar lifetimes.  However, even this top-heavy
IMF is unable to produce enough dust-rich galaxies at $z = 2.5$.  This
suggests that the tension between our fiducial model and high-redshift
observations of massive, dusty galaxies cannot be remedied by a variation in
IMF.

For the fiducial $z = 0.0$ results, we fit data from the L25n512 run with a
Schechter function \citep{Schechter1976} of the form
\begin{equation}
\Phi(M_\text{dust}) \, \Delta M_\text{dust} = \Phi^{*} \left( \frac{M_\text{dust}}{M_\text{dust}^{*}} \right)^{\alpha} \exp\left( \frac{- M_\text{dust}}{ M_\text{dust}^{*}} \right) \, \Delta \left( \frac{M_\text{dust}}{M_\text{dust}^{*}} \right)
\label{EQN:Schechter}
\end{equation}
to determine the best-fitting slope parameter $\alpha$, characteristic dust
mass $M_\text{dust}^{*}$, and normalisation factor $\Phi^{*}$.  We obtain
$\alpha = -1.03$, $M_\text{dust}^{*} = 3.5 \times 10^{6} \, \text{M}_\odot$,
and $\Phi^{*} = 2.2 \times 10^{-2} \, \text{Mpc}^{-3} \, \text{dex}^{-1}$.  For
comparison, the best-fitting Schechter function in \citet{Dunne2011} for $0.0 <
z < 0.1$ produces $\alpha = -1.01$, $M_\text{dust}^{*} = 3.83 \times 10^{7} \,
\text{M}_\odot$, and $\Phi^{*} = 5.87 \times 10^{-3} \, \text{Mpc}^{-3} \,
\text{dex}^{-1}$.  Relative to this observational data, the L25n512 run yields
a similar slope parameter, and though it predicts a lower turnover mass and
higher normalisation factor, Figure~16 in \citet{Dunne2011} demonstrates how
these parameters are degenerate and anti-correlated.

\subsection{Projected Dust Surface Density}

\begin{figure}
\centering
\includegraphics{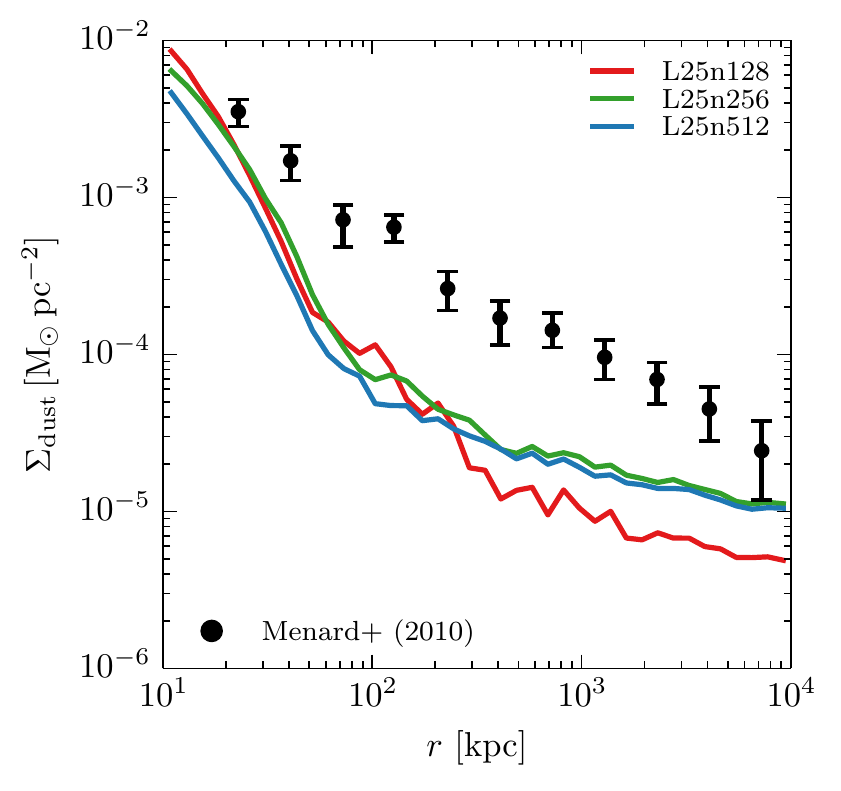}
\caption{Dust surface density~($\Sigma_\text{dust}$) as a function of projected
radius about galactic centres at $z = 0.3$ out to distances of $10 \,
\text{Mpc}$ in physical units.  For each simulation, we show the mean dust
surface density profile averaged across galaxies with $17 < i < 21$, using
projections along the $z$ axis of our box to ensure random orientations.  This
enables comparison with observational data from \citet{Menard2010}, shown in
black, where galaxy position and quasar brightness correlations are used to
infer reddening and SMC-type dust is assumed.  The simulated dust
surface density scaling has lower normalisation than the observed result,
particularly at large radii.}
\label{FIG:dust_surface_density}
\end{figure}

The visualizations in Figure~\ref{FIG:box_evolution} suggest that lines of
sight far from galaxies suffer little dust extinction.  We can directly
quantify this by considering the dust surface density in galactic haloes.  One
observational technique to detect dust in haloes involves cross-correlating the
brightness of quasars with the position of galaxies to infer reddening from
dust \citep{Menard2010}.  This correlation is used to estimate galactic
reddening and infer dust surface density profiles, with the mean dust surface
density following the scaling $\Sigma_\text{dust} \propto r^{-0.8}$.  This
relation has been reproduced by analytic halo models \citep{Masaki2012}, and a
similar technique has been used to study the distribution of dust on larger
scales in galaxy clusters \citep{McGee2010}.

In Figure~\ref{FIG:dust_surface_density}, we show the dust surface density
profile as a function of projected radial distance in physical units around
galactic centres at $z = 0.3$, averaging over all galaxies with $17 < i < 21$
to match the magnitude cut used in \citet{Menard2010}.  We calculate apparent
magnitudes for simulated galaxies using the procedure outlined in Section~3.1
of \citet{Torrey2014}.  Briefly, we use the stellar population synthesis model
of \citet{Bruzual2003} and assign a luminosity to each star particle as a
function of its age, initial stellar mass, and metallicity, and then we set
each galaxy's luminosity to be the sum of the luminosities for constituent star
particles.  To determine apparent magnitudes, we use the luminosity distance
$D_L = 1598 \, \text{Mpc}$ for $z = 0.3$ in our cosmology.  We perform
projections for individual galaxies along the $z$ axis of our simulated box,
resulting in random orientations with respect to the projection axis.  Every
projection is carried out in a cylindrical volume centered on the galactic
potential minimum, using a radius of $10 \, \text{Mpc}$ and a half-height of
$20 \, \text{Mpc}$.  Reducing this cylinder height by a factor of two leaves
the profiles in Figure~\ref{FIG:dust_surface_density} virtually unchanged for
radii less than $1 \, \text{Mpc}$ and lowers them by by about $0.2 \,
\text{dex}$ at the maximum radius of $10 \, \text{Mpc}$.  The mean dust surface
density profile for all galaxies in this magnitude range is the result shown in
Figure~\ref{FIG:dust_surface_density}.  As noted in \citet{Menard2010}, on
scales larger than the virial radius, the dust surface density profile may be
influenced by dust from surrounding or overlapping galaxies.

The simulated dust surface density profiles appear well-converged out to $r
\approx 50 \, \text{kpc}$, with the two highest-resolution runs showing slightly
greater dust surface density out to Mpc scales.  Compared to the observed
$\Sigma_\text{dust} \propto r^{-0.8}$ scaling, the simulated profiles are
steeper for $r < 100 \, \text{kpc}$ and flatter for $r > 100 \, \text{kpc}$.
The dust surface density from our even highest resolution run still lies below
the observed data, with the tension largest for large radial distances.

\subsection{Cosmic Dust Density}\label{SEC:rho_dust}

\begin{figure}
\centering
\includegraphics{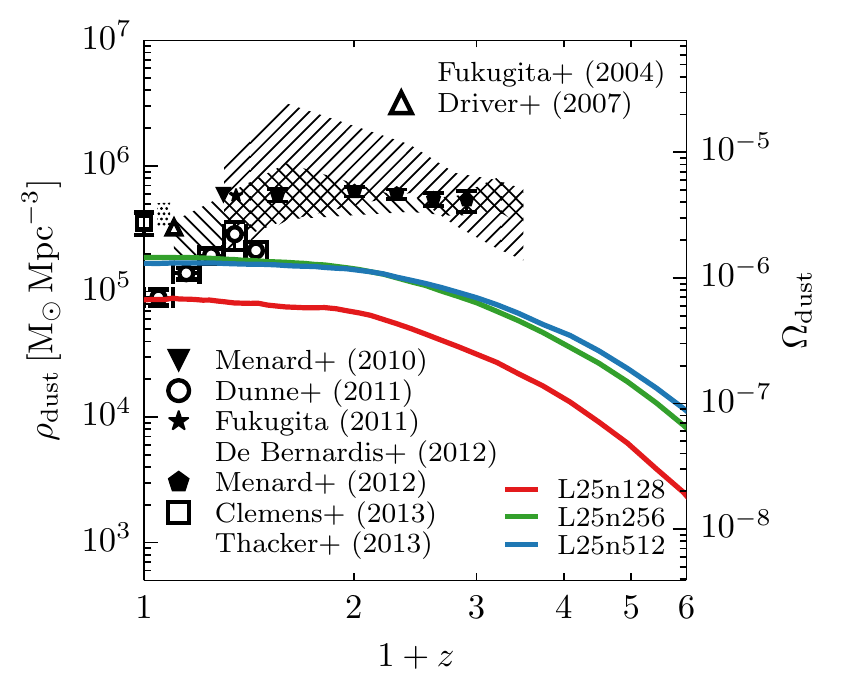}
\caption{Evolution of the comoving cosmic dust density~($\rho_\text{dust}$;
left axis) and associated dust density parameter~($\Omega_\text{dust} =
\rho_\text{dust}/\rho_\text{c}$; right axis) as a function of redshift for our
three resolution simulations~(coloured lines).  Recent observations are shown in
black \citep{Fukugita2004, Driver2007, Menard2010, Dunne2011, Fukugita2011,
DeBernardis2012, Menard2012, Clemens2013, Thacker2013}.  Filled points include
the contribution of dust from haloes, while open points and shaded regions
track only galactic dust.  The $z = 0$ value of $\Omega_\text{dust}
\approx 1.3 \times 10^{-6}$ is similar to observational estimates, though the
simulated cosmic dust density doesn't display the observed decline from $z = 1$
to $z = 0$.}
\label{FIG:integrated_dust_density}
\end{figure}

In Figure~\ref{FIG:integrated_dust_density}, we show the comoving cosmic dust
density $\rho_\text{dust}$ as a function of redshift for our simulations and
compare with observational data at low redshift.  The observational data have
been corrected as in Figure~\ref{FIG:dmf} to conform to the cosmology that we
adopt for our simulations and, where appropriate, the dust mass absorption
coefficient used in \citet{Dunne2011}.  In our simulations, the cosmic dust
density is computed by summing the dust masses of all gas cells and dividing by
the total comoving volume of $(25 \, h^{-1} \, \text{Mpc})^{3}$.  We also show
the cosmic dust density parameter $\Omega_\text{dust} \equiv
\rho_\text{dust}/\rho_\text{c}$.  In our fiducial simulations, the cosmic dust
density increases by over $1 \, \text{dex}$ from $z = 5$ to $z = 0$, although
there is very little evolution for $z < 1.5$ as the cosmic star formation rate
density declines.  Compared to the DMFs presented in
Figure~\ref{FIG:dmf}, the cosmic dust density results presented
Figure~\ref{FIG:integrated_dust_density} show stronger convergence.  In
particular, the L25n256 and L25n512 runs produce $\rho_\text{dust}$ values that
differ by less than $0.1 \, \text{dex}$ for $z < 2$.  Even the low-resolution
L25n128 run displays the same qualitative behaviour with a lower
normalisation.  This suggests that convergence is less of an issue when looking
at volume-integrated dust quantities.

At $z = 0$, the L25n512 run reaches the values $\rho_\text{dust} \approx 2
\times 10^{5} \, \text{M}_\odot \, \text{Mpc}^{-3}$ and
$\Omega_\text{dust} \approx 1.3 \times 10^{-6}$, in rough agreement
with various low-redshift observations.  In comparison, integrating the
best-fitting Schechter function for the L25n512 DMF yields $\Omega_\text{dust}
\approx 6 \times 10^{-7}$ for the dust content of the ISM at $z = 0$.  However,
we do not reproduce the observed decline in $\rho_\text{dust}$ by about a
factor of three from $z \approx 0.35$ to $z = 0$ seen in Herschel ATLAS data
\citep{Dunne2011}.  This decline has have a similar effect on the DMF for this
redshift range, causing a drop in the Schechter function parameter
$M^{*}_\text{dust}$ and shifting the DMF to lower dust masses.  The redshift
behaviour of the cosmic dust density is not as well-studied as those of the
cosmic star formation rate density and stellar mass density \citep[e.g.
see][and references therein]{Madau2014} and would benefit from additional
observations.  We note that observations of the stellar mass density $\rho_{*}$
show an increase of more than $1.5 \, \text{dex}$ from $z = 5$ to $z = 0$ and a
flattening for $z < 1$, results similar to our simulated $\rho_\text{dust}$
evolution.

Observational estimates of $\rho_\text{dust}$ at low redshift have been
obtained in a number of ways.  One method includes fitting a Schechter function
to DMF data and integrating it against dust mass to find $\rho_\text{dust} =
\Gamma(2+\alpha) \, \Phi^{*} \, M_\text{dust}^{*}$, where $\alpha$, $\Phi^{*}$,
and $M_\text{dust}^{*}$ are the best-fitting Schechter parameters
\citep{Dunne2011}.  Others assume a constant ratio between dust mass and
$B$-band luminosity and calculate $\rho_\text{dust}$ by scaling the observed
cosmic luminosity density \citep{Driver2007}.  The integrated dust density can
also be estimated by transforming the luminosity function obtained from
photometric surveys \citep{Clemens2013} or derived from far-infrared power
spectrum measurements \citep{DeBernardis2012, Thacker2013}, or by combining a
constant dust-to-metal ratio, mean ISM metallicity, and cool gas density
parameter \citep{Fukugita2004}.  We note, however, that these calculations tend
to underestimate or neglect dust in galactic haloes, which is thought to
contribute almost as much to $\Omega_\text{dust}$ as ISM dust
\citep{Menard2010, Fukugita2011}.  Dust surface density profiles like in
Figure~\ref{FIG:dust_surface_density}, observationally obtained through
quasar-galaxy reddening measurements \citep{Menard2010, Menard2012}, can be
integrated out to the virial radius to estimate the halo component of dust mass
and in turn a value of $\rho_\text{dust}$ that includes contributions from the
ISM and CGM.

The measurement of $\Omega_\text{dust}$ by \citet{Menard2010} in
Figure~\ref{FIG:integrated_dust_density} accounts for dust in galactic haloes
and lies above other low-redshift observations.  This suggests that
calculations of $\rho_\text{dust}$ and $\Omega_\text{dust}$ using galactic DMFs
tracing ISM luminosity or metallicity data may be underestimating the true
cosmic dust density, especially in cases where galactic outflows can drive dust
away from the ISM.  The results in Figures~\ref{FIG:dust_surface_density}
and~\ref{FIG:integrated_dust_density} also show that $\Sigma_\text{dust}$ and
$\rho_\text{dust}$ are interconnected: the dust content of the ISM cannot be
varied independently of the dust content in galactic haloes, and
$\rho_\text{dust}$ is influenced by dust in both of these regions.  To a large
degree, $\rho_\text{dust}$ determines the normalisation of quantities like
$\Sigma_\text{dust}$ and can be used to put constraints on the typical dust
surface density seen for individual galaxies.

\subsection{Dust on the Star Formation Main Sequence}

\begin{figure*}
\centering
\includegraphics{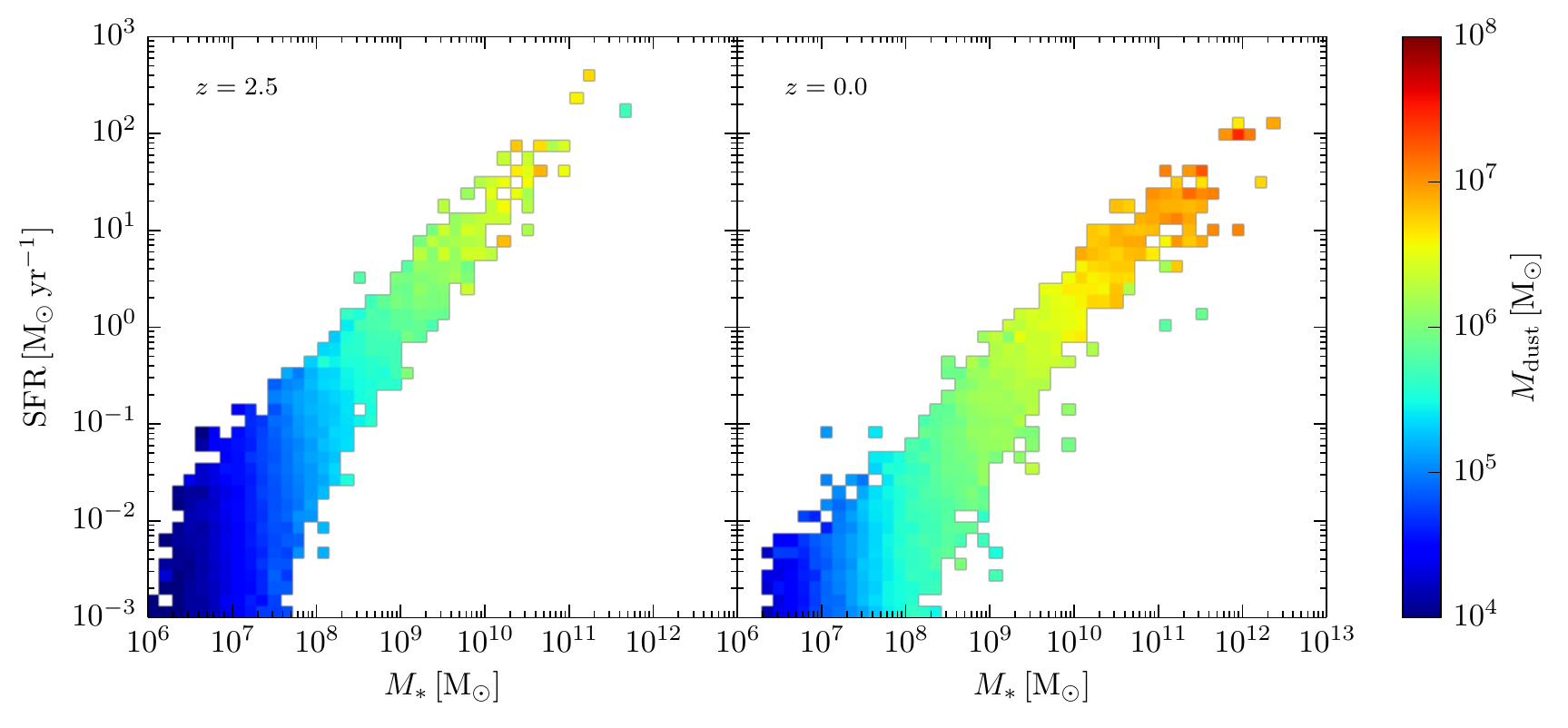}
\caption{Star formation main sequence at $z = 2.5$~(left) and $z = 0.0$~(right)
for our L25n512 run, where the colour of each bin denotes the
average dust mass of galaxies whose stellar mass and SFR fall in those
intervals.  At both redshifts, dust mass tends to increase with stellar mass
and SFR, though the correlation is stronger for stellar mass.}
\label{FIG:star_formation_main_sequence}
\end{figure*}

Figure~\ref{FIG:star_formation_main_sequence} shows two-dimensional histograms
indicating the average dust mass of galaxies on and around the star formation
main sequence at $z = 2.5$ and $0.0$.  The average dust mass tends to increase
with both stellar mass and SFR as seen in both starburst \citep{Magnelli2012,
Santini2014} and local galaxies \citep{Draine2007, Leroy2007, Kennicutt2009,
Galametz2011, Skibba2011, Fisher2013}.  For fixed stellar mass or SFR, average
dust mass increases from $z = 2.5$ to $z = 0.0$, even as the global SFR density
and thus the stellar injection rate of dust drops.

We analyse the dependence of dust mass on stellar mass and SFR using a
least-squares fit to the functional form
\begin{equation}
\log\left( \frac{M_\text{dust}}{M_{\text{dust},0}} \right) = \alpha \log\left( \frac{M_{*}}{10^{10} \, \text{M}_\odot} \right) + \beta \log\left( \frac{\text{SFR}}{\text{M}_\odot \, \text{yr}^{-1}} \right),
\end{equation}
where $M_{\text{dust},0}$, $\alpha$, and $\beta$ are free parameters.  We
apply this fit to all galaxies within $1\sigma$ of the star formation main
sequence, using the best-fitting relations from Figure 7 of \citet{Torrey2014}.
(The $z = 2.0$ main sequence relation in that work is used for our $z = 2.5$
panel.)  We also impose the cut $M_{*} > 10^{7} \, \text{M}_\odot$ to
avoid galaxies that lie at the poorly-resolved end of the galaxy stellar mass
function.  At $z = 2.5$, the best-fitting parameters are $M_{\text{dust},0} =
2.3 \times 10^{6} \, \text{M}_\odot$, $\alpha = 0.55$, and $\beta = 0.11$,
while at $z = 0.0$ they are $M_{\text{dust},0} = 3.7 \times 10^{6} \,
\text{M}_\odot$, $\alpha = 0.43$, $\beta = 0.09$.  Dust mass is largely
predicted by stellar mass, although there is also a weak scaling with SFR.

Figure~\ref{FIG:star_formation_main_sequence} suggests that it is reasonable to
associate the dustiest galaxies with those that are the most star-forming.
Such a procedure is used in some hydrodynamical simulations where dust is not
directly treated in order to study the highly-luminous, dust-rich submillimetre
population.  For example, \citet{Dave2010} assumes a $z \sim 2$ submillimetre
galaxy number density of $1.5 \times 10^{-5} \, \text{Mpc}^{-3}$ based on
observations \citep{Chapman2005, Tacconi2008} and finds that an SFR cut of
around $180 \, \text{M}_\odot \, \text{yr}^{-1}$ produces a galaxy population
with a similar number density.  This highly star-forming population is taken to
be the submillimetre set.

However, we caution that the relation between dust mass, stellar mass, and SFR
is complex, especially for submillimetre galaxies.  Our boxes of side length
$25 \, h^{-1} \, \text{Mpc}$ have difficulty capturing the nuclear starbursts
and main sequence outliers \citep{Sparre2015, Sparre2016} that tend to
simultaneously increase the SFR and decrease the dust mass \citep{Hayward2011}.
Radiative transfer predicts that submillimetre flux scales more strongly with
dust mass than SFR, and Figure~\ref{FIG:star_formation_main_sequence} indicates
dust mass is most strongly predicted by stellar mass.  Comparing a low-stellar
mass starburst with a higher-mass, lower-SFR main sequence galaxy, the latter
may have higher submillimetre flux because its increased dust mass more than
compensates for its smaller SFR.

\subsection{Scaling and Evolution of Dust-to-Stellar Mass Ratio}\label{SEC:dust_to_stellar_ratio}

\begin{table*}
\centering
\caption{Observational references with dust-to-stellar mass ratio data shown in
Figure~\ref{FIG:dust_ssfr}.  We provide an approximate redshift range
corresponding to our cosmology for each sample and list the redshift at
which data is plotted in Figure~\ref{FIG:dust_ssfr}.  In the last column, we
briefly characterize each sample of galaxies and clarify which data we use.
Several references provided already-binned data with uncertainties capturing
scatter about the mean.  For those references that provided quantities on a
galaxy-by-galaxy basis, we binned data ourselves, calculating uncertainties in
log-space to provide symmetric error bars for Figure~\ref{FIG:dust_ssfr}.}
\begin{tabular}{lllll}
\hline
Reference & Abbreviation & Redshift Range & Redshift Panel & Notes \\
\hline
\citet{DaCunha2010} & D10 & $z < 0.22$ & $z = 0.0$ & we bin the sample of galaxies observed in all four IRAS bands \\
\citet{Dunne2011} & D11 & $z < 0.5$ & $z = 0.0$ & we use the mean result for these late-type galaxies \\
\citet{Skibba2011} & S11 & $z < 0.01$ & $z = 0.0$ & we use the mean result for galaxies of all morphological types \\
\citet{Bourne2012} & B12-B & $z < 0.35$ & $z = 0.0$ & already-binned sample of galaxies with blue $g-r$ colour \\
& B12-G & $z < 0.35$ & $z = 0.0$ & already-binned sample of galaxies with green $g-r$ colour \\
& B12-R & $z < 0.35$ & $z = 0.0$ & already-binned sample of galaxies with red $g-r$ colour \\
\citet{Cortese2012} & C12-N & $z < 0.01$ & $z = 0.0$ & already-binned sample of H\,\textsc{i}-normal galaxies \\
& C12-D & $z < 0.01$ & $z = 0.0$ & already-binned sample of H\,\textsc{i}-deficient galaxies \\
\citet{Davies2012} & D12 & $z < 0.01$ & $z = 0.0$ & we bin this sample of bright galaxies \\
\citet{Rowlands2012} & R12 & $z < 0.5$ & $z = 0.0$ & we bin the sample of early-type galaxies \\
\citet{Smith2012} & S12 & $z < 0.01$ & $z = 0.0$ & we bin the sample of early-type galaxies, excluding non-detections \\
\citet{Santini2014} & S14 & $0.6 < z < 1.5$ & $z = 1.0$ & we bin this sample of galaxies \\
& S14 & $1.5 < z < 2.5$ & $z = 2.5$ & we bin this sample of high-redshift galaxies \\
\citet{RemyRuyer2015} & R15 & $z < 0.05$ & $z = 0.0$ & we bin the sample covering a roughly $2 \, \text{dex}$ metallicity range \\
\hline
\end{tabular}
\label{TAB:dust_ssfr}
\end{table*}

\begin{figure*}
\centering
\includegraphics{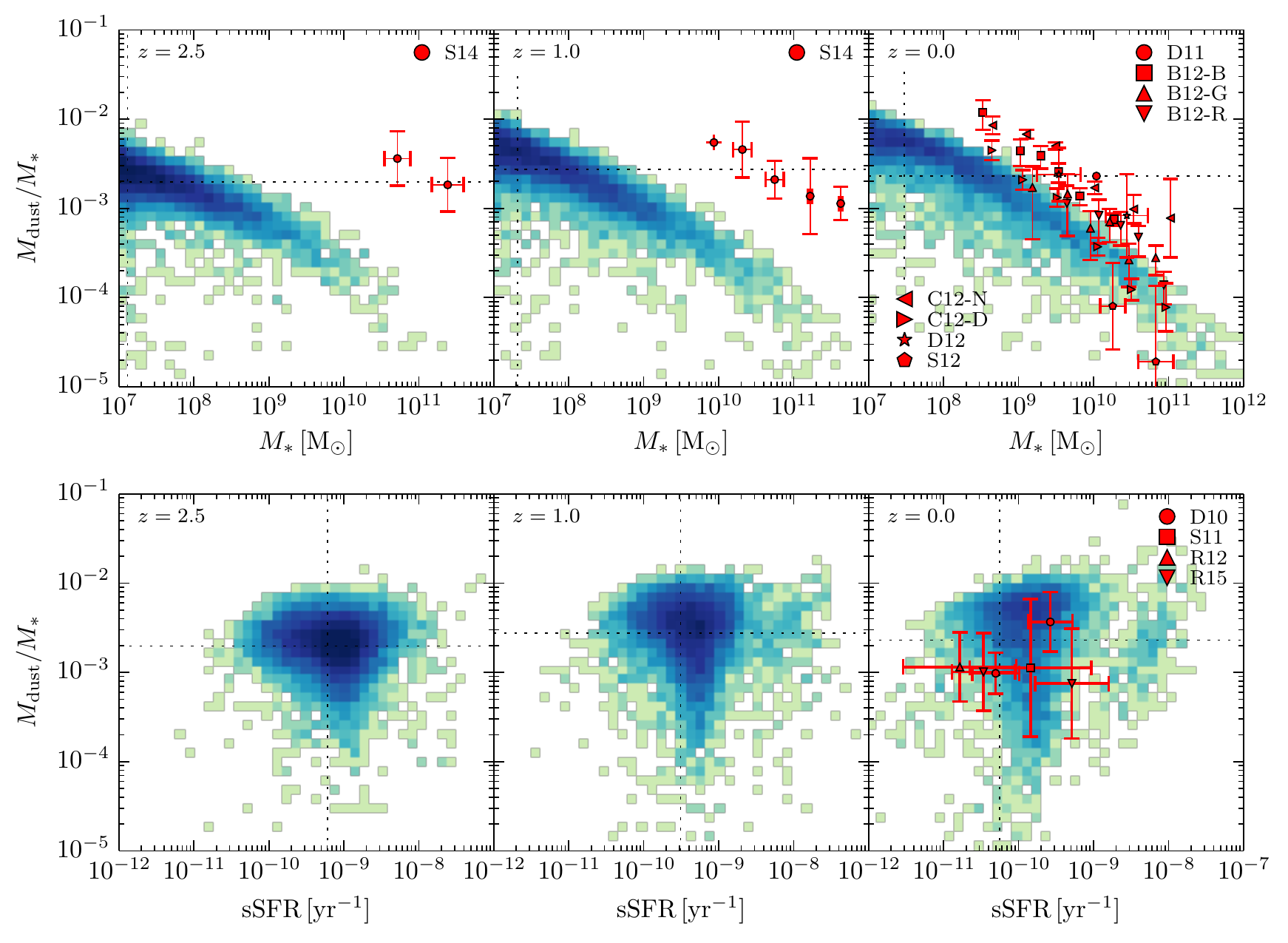}
\caption{Dust-to-stellar mass ratio~($M_\text{dust}/M_{*}$)
as a function of stellar mass (top row) and specific star-formation
rate~(sSFR;~bottom row) at $z = 2.5$, $1.0$, and $0.0$~(left, middle, and
right panels).  For each redshift, the logarithmic distribution of simulated
galaxies in the L25n512 run is given by a two-dimensional histogram, with bluer
colours indicating greater density.  Dotted black lines mark the median value
in the distribution for each axis.  References for the binned observational
data (red points) are given in Table~\ref{TAB:dust_ssfr}.  Dust-to-stellar mass
ratio anti-correlates with stellar mass at both high and low redshift.}
\label{FIG:dust_ssfr}
\end{figure*}

A galaxy's dust-to-stellar mass ratio can increase not only through
dust injected into the ISM during stellar evolution, but also through
subsequent dust growth in collisions with interstellar gas. Chemical evolution
models \citep[e.g.~based on the work in][]{Edmunds2001} suggest that
stellar injection of dust by itself -- with no ISM dust growth -- only
produces dust-to-stellar mass ratios around $10^{-3}$ or less. Even
in the extreme scenario where SNe produce more dust than observed and
condense nearly all ejected metals, this only results in
dust-to-stellar mass ratios near $10^{-2}$ \citep{Dunne2011,
Bourne2012}.  We predict some galaxies with dust-to-stellar mass ratios
around $10^{-2}$, and such dust-to-stellar mass ratios have been observed
\citep{Bourne2012, Cortese2012}.  Unless dust condensation efficiencies are
much higher than expected, this suggests that ISM dust growth is an important
contributor to high dust-to-stellar mass ratios.  We note that previous works
have analyzed the relative strengths of interstellar dust growth and stellar
injection of grains for increasing a galaxy's dust-to-gas ratio
\citep{Mattsson2012, Mattsson2012b}.  By studying a population of galaxies, we
can investigate both galaxies whose dust-to-stellar mass ratios are driven by
stellar injection of grains and by ISM dust growth.

In Figure~\ref{FIG:dust_ssfr}, we show the distribution of our simulated
galaxies as a function of dust-to-stellar mass ratio and stellar mass as well
as dust-to-stellar mass ratio and specific star-formation rate~(sSFR) at $z =
2.5$, $1.0$, and $0.0$.  We compare with multiple sources of observational
data, detailed in Table~\ref{TAB:dust_ssfr} and meant to capture a variety of
morphological types, colours, and metallicities, and note that sSFR
anti-correlates with stellar mass \citep{Brinchmann2004, Salim2007, Karim2011,
Whitaker2012, Abramson2014, Knebe2015}.   Several of the data sets in
Table~\ref{TAB:dust_ssfr} were already binned across stellar mass or sSFR or
provided a mean result which we show, together with quoted uncertainties,
without modification in Figure~\ref{FIG:dust_ssfr}.  For those unbinned data
sets comprised of numerous individual galaxy observations, we manually bin the
data to improve plot readability and compute sample standard deviations in
log-space to obtain symmetric error bars for Figure~\ref{FIG:dust_ssfr}.

The median sSFR drops by over $1 \, \text{dex}$ from $z = 2.5$ to $0.0$ as the
median dust-to-stellar mass ratio is largely unchanged.  The scatter in
dust-to-stellar mass ratio at fixed sSFR slightly increases towards low
redshift.  However, the slope of the dust-to-stellar mass ratio versus stellar
mass relation does not change appreciably from $z = 2.5$ to $z = 0.0$.  At high
redshift, the observed dust-to-stellar mass ratios for $M_{*} \gtrsim 10^{10}
\, \text{M}_\odot$ from \citet{Santini2014} are larger than what we predict.
However, we note that most of the high-redshift galaxies in \citet{Santini2014}
have $\text{SFR} \gtrsim 100 \, \text{M}_\odot \, \text{yr}^{-1}$, making them
more star-forming than nearly all galaxies in our simulation.

\begin{figure}
\centering
\includegraphics{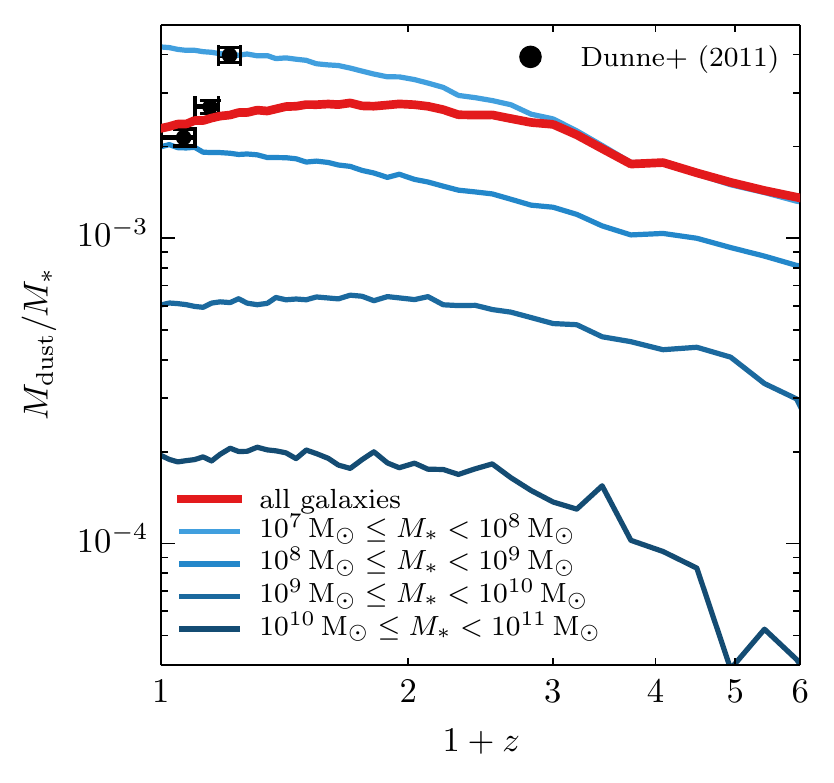}
\caption{Evolution of the galactic dust-to-stellar mass ratio~($M_\text{dust} /
M_{*}$) as a function of redshift for the L25n512 run.  We compute the median
dust-to-stellar mass ratio for all galaxies~(red line) and for different
stellar mass bins~(blue lines, with deeper shades indicating greater stellar
mass), excluding galaxies with no stellar component.  Observational data using
galaxies with spectroscopic redshifts from the Herschel ATLAS~\citep{Dunne2011}
are shown in black, though we note these observations were not binned by mass.
The median dust-to-stellar mass ratio increases by about a factor of
1.7 from $z = 5$ to $z = 0$, with the most massive galaxies displaying
less evolution in dust-to-stellar mass ratio.}
\label{FIG:Mdust_Mstar_evolution}
\end{figure}

To better understand the connection between dust-to-stellar mass ratio and
stellar mass, in Figure~\ref{FIG:Mdust_Mstar_evolution} we plot the median
dust-to-stellar mass ratio as a function of redshift for $1 \, \text{dex}$
stellar mass bins ranging from $10^{7} \, \text{M}_\odot$ to $10^{11} \,
\text{M}_\odot$, along with the median ratio across all simulated galaxies.
The results from Figure~\ref{FIG:dust_ssfr} -- that the dust-to-stellar mass
ratio decreases with increasing stellar mass, while the overall median
dust-to-stellar mass ratio does not substantially increase with time -- are
confirmed in Figure~\ref{FIG:Mdust_Mstar_evolution}.  The nature of the galaxy
stellar mass function implies that at nearly every redshift the median
dust-to-stellar mass ratio lies nearer to the ratio for $10^{7} \,
\text{M}_\odot < M_{*} < 10^{9} \, \text{M}_\odot$ galaxies than for more
massive galaxies.

The overall median dust-to-stellar mass ratio increases by just under a factor
of two from $z = 5$ to $0$, and it is largely flat for $z < 1$.  Low stellar
mass galaxies display similar behaviour, while the dust-to-stellar mass ratio
for large galaxies with $M_{*} > 10^{10} \, \text{M}_\odot$ is roughly
$1 \, \text{dex}$ below the value for galaxies with $10^{7} \, \text{M}_\odot <
M_{*} < 10^{9} \, \text{M}_\odot$.  However, we do not capture the decrease in
dust-to-stellar mass ratio by a factor of two observed in Herschel ATLAS data
for $z < 0.5$ \citep{Dunne2011}, a result similarly shown in
Figure~\ref{FIG:integrated_dust_density}.

\begin{figure*}
\centering
\includegraphics{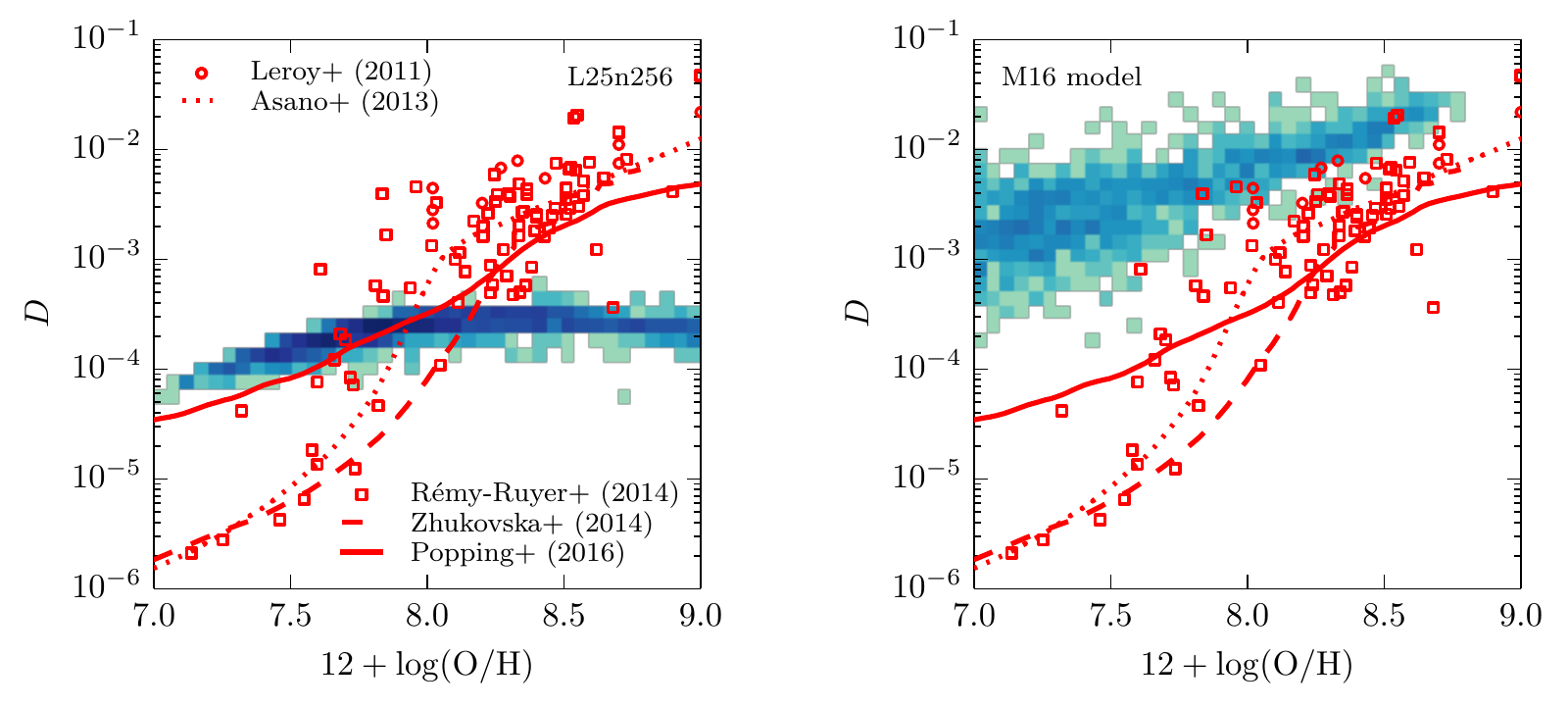}
\caption{Simulated dust-metallicity relations at $z = 0$ for the
fiducial L25n256 run (left) and M16 model (right).  The two-dimensional
histograms use a logarithmic colourscale to indicate number density, with bluer
colours denoting a greater number of galaxies at a given dust-to-gas ratio and
metallicity.  Red points and lines denote observational data \citep{Leroy2011,
RemyRuyer2014} and results from analytic and semi-analytic modelling
\citep{Asano2013a, Zhukovska2014, Popping2016}.  To improve readability, we
omit error bars.  From \citet{Asano2013a} we use the $\tau_\text{SF} = 5 \,
\text{Gyr}$ model and from \citet{Zhukovska2014} we use the ``6x 500 Myr bursts
$\tau_\text{SF} = 2 \, \text{Gyr}$'' model, which were compiled by
\citet{RemyRuyer2014}.}
\label{FIG:dust_metallicity}
\end{figure*}

While the dust-to-stellar mass scaling at $z = 0$ is in decent
agreement with observations, the relation between dust-to-gas ratio and
gas-phase metallicity displays more tension and a greater sensitivity to the
parameters of our model.  In Figure~\ref{FIG:dust_metallicity}, we plot the
dust-metallicity relation at $z = 0$ for the L25n256 and M16 model simulations
and offer a comparison to observations \citep{Leroy2011, RemyRuyer2014} and
modelling \citep{Asano2013a, Zhukovska2014, Popping2016}.  It is clear that the
L25n256 run fails to match slope of the expected dust-metallicity relation:
despite offering a reasonable fit to the $z = 0$ DMF in
Figure~\ref{FIG:dmf}, the dust-metallicity relation is far too flat.  The
L25n256 run is similar to the semi-analytic model of \citet{Popping2016} for
$12 + \log(\text{O}/\text{H}) < 8$, but deviates strongly at high metallicity.
The dust growth timescale in large galaxies in the L25n256 run seems to be too
long, preventing them from rapidly growing their dust mass.  On the other hand,
the M16 model -- which employs a stronger ISM dust growth mechanism and lacks
thermal sputtering -- displays a dust-metallicity relation whose slope better
matches observations but whose normalisation is too high.  However, the M16
model significantly overproduces dust-rich galaxies in its $z = 0$ DMF.
Figure~\ref{FIG:dust_metallicity} offers another look at the DMF tension in
Figure~\ref{FIG:dmf} and highlights the difficulty in producing enough dust to
match the dust-metallicity relation while avoiding a DMF with too many
dust-rich galaxies.

\subsection{Stellar Population Synthesis Postprocessing}

We can combine the direct dust mass tracking in our work with stellar
population synthesis postprocessing to make predictions about the observational
properties of simulated galaxies.  One such property is the dust-to-stellar flux
ratio ($f_\text{dust}/f_{*}$), which measures the flux reradiated by dust grains
as a fraction of unextincted stellar flux and has been observed to correlate
with dust mass and infrared luminosity \citep{Skibba2011}.  Below, we use the
\textsc{fsps} \citep{Conroy2009, Conroy2010} stellar population synthesis code
to estimate each galaxy's bolometric luminosity and in turn its dust-to-stellar
flux ratio.

The dependence of optical depth on host galaxy properties was previously
modelled in \citet{Jonsson2006b}.  Following Table~2 in \citet{Jonsson2006b},
we estimate the bolometric attenuation to be
\begin{equation}
\tau = 0.93 \left( \frac{Z}{0.02} \right)^{1.10} \left( \frac{\text{SFR}}{\text{M}_\odot \, \text{yr}^{-1}} \right)^{0.61} \left( \frac{M_\text{b}}{10^{11} \, \text{M}_\odot} \right)^{-0.68},
\label{EQN:tauISM}
\end{equation}
where $M_\text{b}$ is the galaxy's total baryon mass.  We calculate $\tau$
directly on a galaxy-by-galaxy basis using our \textsc{subfind} output.  We
compute an unattenuated synthetic spectral energy distribution (SED) for every
star particle as a function of its initial mass, age, and metallicity, assuming
a \citet{Chabrier2003} IMF, and define a galaxy's SED to be the sum of those
from constituent star particles.  For a galaxy with bolometric luminosity $L$
and optical depth $\tau$ computed in Equation~(\ref{EQN:tauISM}), we calculate
the dust luminosity $L_\text{dust}$ using Equation~6 in
\citet{Jonsson2006b},
\begin{equation}
L_\text{dust} / L = 1 - (1/\tau)(1 - e^{-\tau}).
\end{equation}
The stellar luminosity is then $L_{*} = L - L_\text{dust}$.  In essence,
stellar flux is computed by integrating the attenuated stellar SED, and dust
flux is obtained by integrating over the difference between the unattenuated
and attenuated stellar SEDs.  This calculation of the dust-to-stellar flux
ratio is simpler than estimating and removing stellar emission in the
mid-infrared from a dust-extincted SED \citep{Draine2007, MunozMateos2009,
Skibba2011}, although only possible when postprocessing simulated galaxies.

\begin{figure}
\centering
\includegraphics{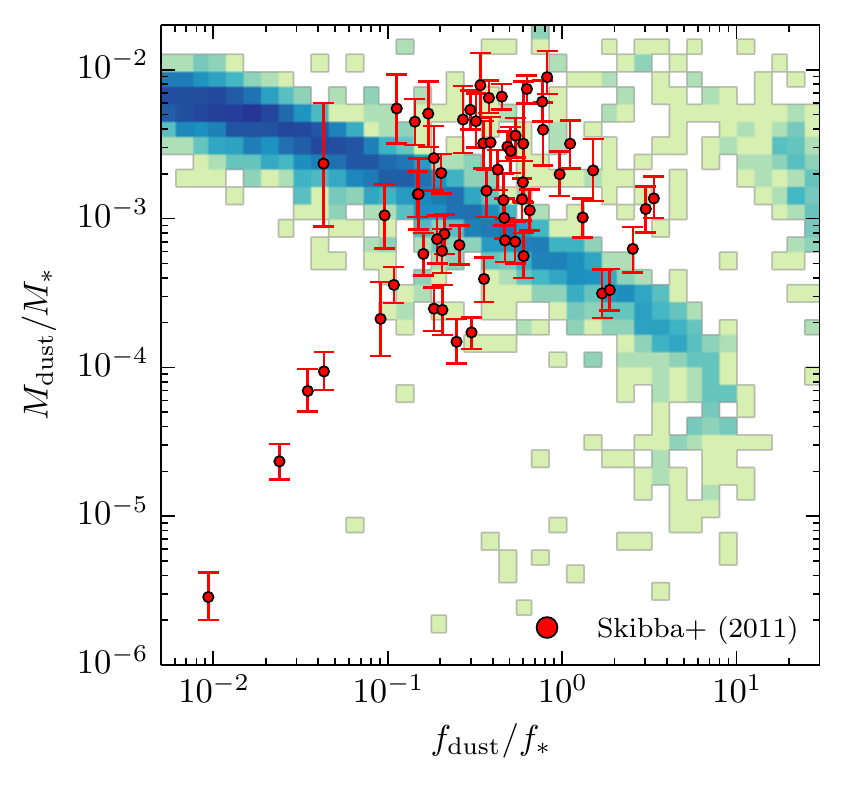}
\caption{Dust-to-stellar mass ratio ($M_\text{dust}/M_{*}$) as a function of
dust-to-stellar flux ratio ($f_\text{dust}/f_{*}$) at $z = 0$.  The
two-dimensional histogram shows the number distribution of galaxies from the
L25n512 run on a logarithmic scale, with bluer colours denoting greater counts.
We compare with observational data (red circles) from the Herschel KINGFISH
Survey \citep{Skibba2011}.}
\label{FIG:flux_ratio}
\end{figure}

Figure~\ref{FIG:flux_ratio} shows the distribution of galaxies in the L25n512
run as a function of dust-to-stellar flux and mass ratios at $z = 0$, with
observational data from the Herschel KINGFISH Survey overlaid
\citep{Skibba2011}.  While the range of dust-to-stellar mass ratios does tend
to match these observations, the simulated dust-to-stellar flux ratios are
biased to smaller values than in the Herschel KINGFISH set.  The
two-dimensional distribution of simulated galaxies peaks near
$M_\text{dust}/M_{*} \approx 5 \times 10^{-3}$ and $f_\text{dust}/f_{*} \approx
2 \times 10^{-2}$, though there is significant scatter in the dust-to-stellar
flux ratios, with a number of galaxies recording $f_\text{dust}/f_{*} > 1$.
The largest tension with observations comes at low dust-to-stellar flux ratio,
where we predict numerous galaxies with large dust-to-stellar mass ratio.  From
Figure~\ref{FIG:dust_ssfr}, we know that galaxies with high dust-to-stellar
mass ratios tend to have low stellar masses and thus low metallicities
and SFRs.  This may drive down the optical depths calculated in
Equation~(\ref{EQN:tauISM}) and thus the dust-to-stellar flux ratios.  Several
of the galaxies in \citet{Skibba2011} with low dust-to-stellar flux ratios are
early-types, which are known to have smaller dust-to-stellar mass ratios on
average than spirals \citep{Rowlands2012, Smith2012}.  In any case, the scatter
in our simulated results does confirm the observation that the dust-to-stellar
flux ratio covers roughly three orders of magnitude and does not effectively
constrain the mass ratio \citep{Skibba2011}.

Previous works have demonstrated how hydrodynamical simulations without direct
dust tracking can be coupled with radiative transfer to study galactic flux
densities at submillimetre wavelengths \citep{Chakrabarti2008, Narayanan2009,
Narayanan2010, Hayward2011, Hayward2012, Hayward2013}.  Performing dust
radiative transfer on simulations of isolated and merging disc galaxies,
\citet{Hayward2011} developed fitting functions to estimate submillimetre flux
densities in the SCUBA $850 \, \mu\text{m}$ and AzTEC $1.1 \, \text{mm}$ bands
as a function of SFR and dust mass as well as dust luminosity and dust
mass.  While these relations were derived from simulations investigating number
counts of bright submillimetre galaxies, here we apply them to our full sample
of galaxies to demonstrate how cosmological simulations can benefit from
results obtained through radiative transfer calculations.

To construct submillimetre number densities, we define the Hubble
parameter $H(z) = H_0 \sqrt{\Omega_\text{m}(1+z)^3 + \Omega_\Lambda}$ and
comoving distance
\begin{equation}
l_\text{c}(z) = \int_{0}^{z} \frac{c \diff z'}{H(z')}.
\label{EQN:comoving_distance}
\end{equation}
If $\Phi(S, z)$ denotes the comoving number density of galaxies with
submillimetre flux $S$ at redshift $z$ per unit logarithmic flux, then
\begin{equation}
\phi(S) = \left( \frac{\pi}{360^2} \, \text{deg}^{-2} \right) \int_{0}^{\infty} 4 \pi l_\text{c}(z)^2 \Phi(S, z) \diff l_\text{c}(z)
\end{equation}
is the number of galaxies with flux $S$ per square degree per unit logarithmic
flux.  Simplifying, we calculate submillimetre number counts using
\begin{equation}
\phi(S) = \left( \frac{\pi}{360^2} \, \text{deg}^{-2} \right) \int_{0}^{\infty} 4 \pi l_\text{c}(z)^2 \Phi(S, z) \frac{c}{H(z)} \diff z.
\label{EQN:phi_S_integral}
\end{equation}
In practice, we numerically integrate Equation~(\ref{EQN:phi_S_integral}) using
$\Phi(S, z)$ values constructed from simulation output at
discrete redshifts.

\begin{figure}
\centering
\includegraphics{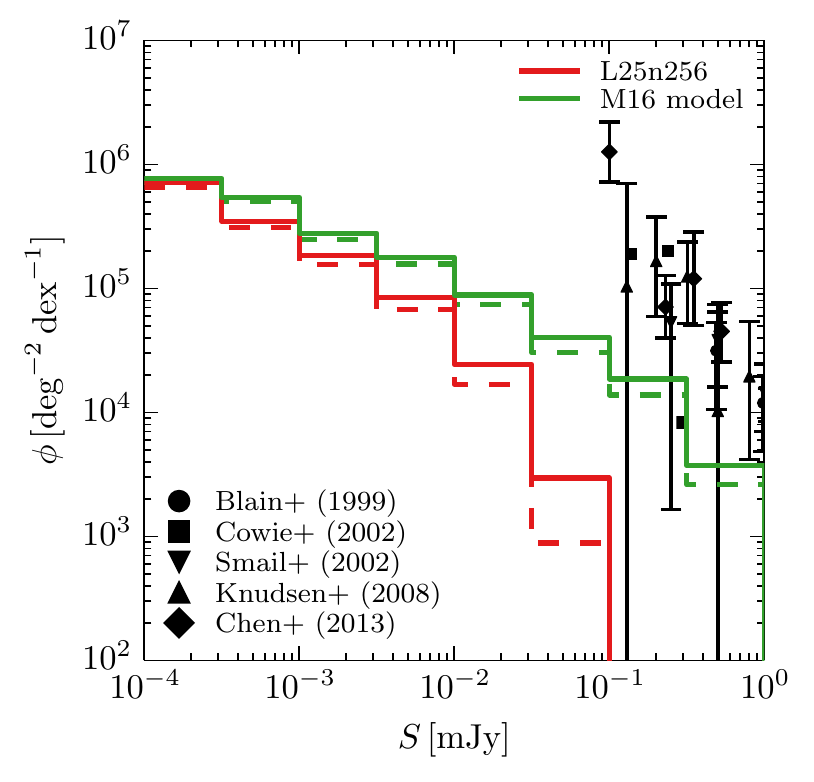}
\caption{Simulated number density functions in the SCUBA $850 \,
\mu\text{m}$ band for the fiducial L25n256 run (red)~and the M16
model~(green).  Fluxes are computed using the luminosity- and dust
mass-dependent fitting functions provided in \citet{Hayward2011}.
Black points mark observations for $850 \, \mu\text{m}$
\citep{Blain1999b, Smail2002, Cowie2002, Knudsen2008, Chen2013} as compiled by
\citet{Casey2014}.  Because the fits in \citet{Hayward2011} were not designed
for $z \lesssim 1$, we show two versions of the number density functions: one
integrated all the way down to $z = 0$ (solid lines) and the other only down to
$z = 1$ (dashed lines).  The versions show similar behaviour.  While the M16
model offers a better fit to the observed submillimetre number densities and
the high-redshift DMF in Figure~\ref{FIG:dmf}, it significantly overproduces
dust in the $z = 0$ DMF.}
\label{FIG:flux_density}
\end{figure}

Number density functions for simulated galaxies at $850 \,
\mu\text{m}$ are shown in Figure~\ref{FIG:flux_density} for our fiducial
L25n256 run and the M16 model.  We compare with various observational
data \citep{Blain1999b, Smail2002, Cowie2002, Knudsen2008, Chen2013}.  Fluxes
in this band are computed using dust luminosities and dust masses
following Equation~2 and Appendix~A in \citet{Hayward2011}.  Dust
luminosities are calculated as in the construction of
Figure~\ref{FIG:flux_ratio}.  Because the fits provided in
\citet{Hayward2011} are not designed to apply to $z \lesssim 1$, we compute
number density functions in two ways: one using $z = 0$ as the lower limit of
the integration in Equation~(\ref{EQN:phi_S_integral}) and the other using $z =
1$.  This variation changes results only slightly. Results for this
submillimetre band show that the number density of galaxies declines
over the flux interval $10^{-4} \, \text{mJy} < S < 10^{-1} \,
\text{mJy}$ accessible in both simulations.  The M16 model offers a
much better fit to the high-flux observations than the L25n256 run, in large
part because the high-redshift DMF for M16 model in Figure~\ref{FIG:dmf}
contains many more dust-rich galaxies.  However, despite the more realistic
submillimetre number counts, the M16 model has tension of its own: its $z = 0$
DMF contains far too many galaxies with $M_\text{dust} > 10^{8} \,
\text{M}_\odot$.  This tension highlights the need to form enough dust at high
redshift to generate realistic submillimetre number counts while preventing an
excess of dust at low redshift. Futhermore, as noted in our
discussion of the star formation main sequence in
Figure~\ref{FIG:star_formation_main_sequence}, the box length of $L = 25 \,
h^{-1} \, \text{Mpc}$ used in this work makes it difficult to truly probe the
submillimetre regime and uncover possible exponential cutoffs in the
submillimetre number density functions.  This may be worth pursuing in
cosmological simulations of larger volumes or in semi-analytic models with dust
tracking.

\section{Discussion}\label{SEC:discussion}

We have presented full volume cosmological simulations with a model for dust
production and destruction to study the coevolution of dust and galaxies.
Our model offers rough agreement with low-redshift observations of the DMF,
cosmic dust density, and relation between dust-to-stellar mass ratio and
stellar mass, but it also highlights limitations that appear more fundamental.
Despite offering a reasonable match to the $z = 0$ DMF, the fiducial model
fails to capture the abundance of dusty galaxies at high redshift and instead
produces galaxies whose dust masses grow roughly in a monotonic fashion.  It
has been suggested that perhaps the dust-rich galaxies at high redshift have
extra-high star-formation efficiencies, are more efficient at forming dust from
stars, or feature more top-heavy IMFs than low-redshift galaxies
\citep{Dunne2011}.  In this scenario, the most dusty galaxies at $z = 2.5$
could evolve to the present with much lower dust masses after consuming their
gas and dust in star formation.  The ability of the fiducial model to roughly
match the $z = 0$ DMF but not capture the decline in dusty galaxies from high
to low redshift suggests that dust evolution processes may be more dependent on
host galaxy properties like SFR or gas fraction than assumed in this work.  For
example, dust yields in stellar ejecta may be a function of local ISM density
or temperature, which evolve with redshift.  To account for the observed shift
towards lower masses in the DMF from $z = 2.5$ to $z = 0$, we need efficient
sputtering of grains in dust-rich galaxies.  This would better enable different
galaxies to have diverse dust mass histories and perhaps lead to DMF behaviour
that more closely follows the cosmic SFR evolution \citep[e.g. see][and
references therein]{Madau2014}.

Our analysis of the DMF, dust surface density profiles, and cosmic dust density
evolution also highlights the observational uncertainties that make it
challenging to obtain reliable dust mass estimates.  For example, the dust
surface density comparison in Figure~\ref{FIG:dust_surface_density} shows that
our simulated dust radial profiles are closest to observations from
SDSS out to $r \approx 30 \, \text{kpc}$ but lie below observations by
up to $1 \, \text{dex}$ at larger radii.  The cosmic dust density in
our simulations could easily absorb a factor of two or three increase and still
be consistent with observations in Figure~\ref{FIG:integrated_dust_density},
and such a normalisation change would help boost the dust surface density
profiles on a global scale.  This change is plausible since dust condensation
efficiencies in stellar ejecta and ISM growth time-scales are not
well-constrained.  Independent of such a normalisation shift, it is also
possible that thermal sputtering is slightly too strong or galactic outflows
too weak, limiting the amount of dust in galactic haloes.

The sample of SDSS galaxies used to construct surface density profiles in
\citet{Menard2010} has a redshift distribution that peaks at $z \approx 0.3$ but
with a full-width at half-maximum of $0.4$.  While we do not predict much
evolution in the cosmic dust density for $z \lesssim 1$, the fact that
Figure~\ref{FIG:dust_surface_density} is constructed at a fixed redshift of $z
\approx 0.3$ may introduce some deviation in surface density profiles from the
observed result.  The observed reddening signal was also tested for possible
systematic effects (e.g.~by subsampling quasars according to magnitude bins or
using sky regions with different Galactic reddening) and found to be robust.
The calculations in \citet{Menard2010} assume SMC-type dust, motivated in part
by the observation that few high-redshift galaxies share the $0.2 \,
\mu\text{m}$ extinction curve bump characteristic of the Milky Way, but
adopting Milky Way-type dust would change dust masses about a factor of two.
Given the difficulties in estimating dust masses from reddening signals,
including weak constraints on dust mass absorption coefficients, the
discrepancies between simulated and observed dust surface density profiles
could be influenced by inaccuracies in modelled physics like thermal
sputtering or galactic outflows or by uncertain observational assumptions.
The \citet{Menard2010} relation can possibly be used as a
constraint on outflow physics, as the dust surface density profiles at large
radii are likely sensitive to the outflow model and the coupling of dust and
gas in winds.  In our simulations, we assume winds have the same depletion as
the ISM from which they are launched (e.g.~if a wind particle is created in a
cell where 10\% of metals are locked in dust, then 10\% of the metal content of
the wind particle is assumed to be dust), but alternative models that couple
dust more strongly to winds may help drive dust to Mpc distances.  Simulations
by \citet{Zu2011} suggest that reproducing the \citet{Menard2010} relation
without galactic winds is difficult, and the strength of outflows could be used
to constrain enrichment in the intergalactic medium.

While dust masses can be hard to estimate, the findings in
Section~\ref{SEC:dust_to_stellar_ratio} demonstrate the connection between a
galaxy's stellar mass and its dust content.  For example,
Figure~\ref{FIG:Mdust_Mstar_evolution} provides a method to calculate a
galactic dust mass when only the stellar mass and redshift are known, using the
dust-to-stellar mass ratio in the appropriate stellar mass bin.  Both
Figures~\ref{FIG:dust_ssfr} and~\ref{FIG:Mdust_Mstar_evolution} highlight the
dependence of dust-to-stellar mass ratio on stellar mass and why assuming a
uniform dust-to-stellar mass ratio is not ideal.  Previous observational
studies have shown that gas fraction and molecular gas fraction decrease with
stellar mass \citep{Leroy2008, Daddi2010, Geach2011, Saintonge2011,
Popping2012, Bauermeister2013, Tacconi2013, Boselli2014, Bothwell2014,
MorokumaMatsui2015}, and this result has been reproduced in galaxy formation
simulations and semi-analytic models \citep{Hopkins2009, Obreschkow2009,
Dave2010, Lagos2011a, Lagos2011b, Duffy2012, Fu2012, Genel2014, Popping2014,
Lagos2015, Narayanan2015}.  Additionally, simple dust and chemical evolution
models suggest that the dust-to-stellar mass ratio increases with gas fraction
\citep{Dunne2011}, a result that has been seen in Herschel Reference Survey
data \citep{Cortese2012}.

Together, these findings indicate the dust-to-stellar mass ratio should be
largest in low stellar mass systems, a result that agrees with the negative
slope of dust-to-stellar mass ratio versus stellar mass shown in
Figures~\ref{FIG:dust_ssfr} and~\ref{FIG:Mdust_Mstar_evolution}.  These less
massive galaxies have high sSFRs that allow the injection of dust from stellar
sources more quickly than in larger systems, and since their sSFRs peak later
than those of more massive galaxies \citep{Cowie1996}, smaller galaxies can see
more growth in the dust-to-stellar mass ratio.  Knowing a galaxy's stellar
mass, and its star-formation history, allows us to better estimate its dust
content.

\section{Conclusions}\label{SEC:conclusions}

In this work, we extended the dust model in the moving-mesh code \textsc{arepo}
to account for thermal sputtering of grains and performed cosmological
hydrodynamical simulations to analyse the evolution of dust in a diverse sample
of galaxies.  We studied the evolution of the DMF, the radial distribution of
dust in galactic haloes and on Mpc scales, and the contribution of dust to the
cosmic mass budget.  Also, we explored how a galaxy's SFR and stellar mass
impacts its dust content.  Our main conclusions are as follows:

\begin{enumerate}
\item Our model broadly reproduces the observed $z = 0$ DMF over the range of
masses accessible in a $(25 \, h^{-1} \, \text{Mpc})^{3}$ volume.  The DMF is
presented for simulations at three resolutions, with the highest-resolution
simulation softening $z = 0$ gravitational forces on scales of $625 \, h^{-1}
\, \text{pc}$.

\item The mean dust surface density profile for simulated galaxies with $17 < i
< 21$ at $z = 0.3$ declines with radial distance, similar to the
$\Sigma_\text{dust} \propto r^{-0.8}$ scaling seen in SDSS data out to
projected distances of $10 \, \text{Mpc}$, although the normalisation of the
simulated dust surface density lies up to $1 \, \text{dex}$ below
observations for $r \gtrsim 100 \, \text{kpc}$.

\item The cosmic dust density parameter at $z = 0$ is estimated to be
$\Omega_\text{dust} = 1.3 \times 10^{-6}$, close to values obtained
from low-redshift observations.  We see little evolution in
$\Omega_\text{dust}$ for $z \lesssim 1.5$, in tension with power
spectrum-derived measurements that show a decline of roughly $0.5 \,
\text{dex}$.  This conflict is consistent with our model's underproduction of
dusty galaxies for the high-redshift DMF.

\item At both high and low redshift, dust mass increases with stellar mass
along the star formation main sequence.  This suggests that semi-analytic or
galaxy formation models without dust tracking can estimate dust content using
the star formation main sequence.  Semi-analytic models may also benefit from
fitting functions for submillimetre number densities.

\item The dust-to-stellar mass ratio is predicted to anti-correlate with
stellar mass at high and low redshift, and this relation parallels
observations at $z = 0$.  Less massive systems witness growth in
dust-to-stellar mass ratio over $0 < z < 5$.  Our simulated galaxies also agree
well with the observed distribution of dust-to-stellar mass ratio versus sSFR
at $z = 0$.

\item By combining direct dust mass tracking with stellar population synthesis
postprocessing, we predict dust-to-stellar mass and flux ratios for our
simulated galaxies at $z = 0$ and compare to observations.  Coupling
with empirical relations from radiative transfer simulations, we estimate the
high-redshift submillimetre number density functions for our sample of galaxies
at $850 \, \mu\text{m}$.

\item While our model reproduces the observed $z = 0$ DMF fairly well, it is
unable to capture the abundance of dust-rich galaxies at high redshift.
Instead, the simulated DMF evolves in a fairly monotonic fashion.  Adopting a
top-heavy IMF does increase the abundance of high-redshift dusty galaxies but
not to the extent seen in observations.

\item To better match the observed DMF evolution, we may need physical
prescriptions that are more closely connected to the behaviour of the cosmic
SFR density and produce more dusty galaxies near the peak of star formation.
For example, adopting non-constant dust condensation efficiencies that vary
with ISM density and temperature may allow the largest galaxies to more
efficiently produce dust at high redshift but limit dust formation at lower
redshifts where star formation is less efficient and the DMF shifts towards
lower masses.
\end{enumerate}

The dust model presented in this work yields low-redshift results in rough
agreement with a number of observables across a diverse sample of galaxies, but
it also highlights areas of tension.  In particular, this model fails to
predict the abundance of dust-rich galaxies at high redshift and the slight
decline in $\Omega_\text{dust}$ as galaxies evolve towards low redshift.
Furthermore, to truly probe the high-redshift submillimetre regime and the
massive end of the DMF will require larger cosmological volumes.  Nonetheless,
this work demonstrates how simulations of large galaxy populations can be used
to study the evolution of dust across diverse environments and the distribution
of dust on cosmological scales.

\section*{Acknowledgements}

We thank Volker Springel for providing us with access to \textsc{arepo}.
We also thank the referee for their constructive feedback.

The simulations were performed on the joint MIT-Harvard computing cluster
supported by MKI and FAS.  RM acknowledges support from the DOE CSGF under
grant number DE-FG02-97ER25308.  MV acknowledges support through an MIT RSC
award.  CCH is grateful to the Gordon and Betty Moore Foundation for financial
support.

\bibliographystyle{mn2e}
\bibliography{../bibliography}

\appendix

\section{Analytical Thermal Sputtering Calculations}\label{SEC:appendix_sputtering}

The empirical thermal sputtering rate given in Equation~(\ref{EQN:adot})
falls off quickly for $T \lesssim 10^{6} \, \text{K}$.  In this section, we
detail how such a temperature dependence arises from analytical calculations of
collisions between gas atoms and grains and the subsequent erosion of grains
from sputtering.  We refer the reader to the existing literature
\citep{Barlow1978, Draine1979b, Tielens1994} for more thorough analysis.

Following Equations 4.19 and 4.20 in \citet{Tielens1994}, consider a grain of
radius $a$ in a medium of temperature $T$.  Then, the number of particles
sputtered off of the grain surface per unit time is given by
\begin{equation}
\frac{\diff N_\text{sp}}{\diff t} = 2 \pi a^2 \sum_{i} n_i \langle Y_i v \rangle,
\end{equation}
where the leading factor of two accounts for collisions at non-normal angles,
$\pi a^2$ is the grain cross-section, and $n_i$ and $\langle Y_i v \rangle$ are
the number density and Maxwell-Boltzmann distribution-averaged product of
sputtering yield and velocity for a gas ion of species $i$.  Here, $Y_i$
measures the number of particles sputtered from the grain per gas ion collision
\citep[e.g.~studied in detail in Section~4.1 of][]{Tielens1994}, and the
Maxwell-Boltzmann distribution corresponds to temperature $T$.

Suppose the grain has mass $m$ and uniform internal density $\rho_\text{g}$ and
that particles sputtered from the grain surface have mass $m_\text{sp}$.  For
example, a carbonaceous grain might have $m_\text{sp} = m_\text{C}$, the mass
of a carbon atom.  The grain mass loss rate
\begin{equation}
\frac{\diff m}{\diff t} = 4 \pi a^2 \rho_\text{g} \frac{\diff a}{\diff t}
\end{equation}
implies that the change in grain radius per unit time due to thermal sputtering
is given by
\begin{equation}
\frac{\diff a}{\diff t} = \frac{n_\text{H} m_\text{sp}}{2 \rho_\text{g}} \sum_{i} A_i \langle Y_i v \rangle,
\end{equation}
where $A_i$ is the abundance of gas ions of species $i$.  This sputtering rate
is a function of temperature due to its averaging over a Maxwell-Boltzmann
distribution.  Combined with analytic models of sputtering yields, the thermal
sputtering rate shows a sharp drop-off for $T \lesssim 10^{6} \, \text{K}$.
Thus, the empirical formula given by Equation~(\ref{EQN:adot}) captures the
essential temperature dependence and normalisation of the thermal sputtering
rate and avoids the need to calculate Maxwell-Boltzmann distribution-averaged
sputtering integrals in our simulation code.

\section{Variation of Grain Size Parameter}\label{SEC:appendix_grain}

\begin{figure}
\centering
\includegraphics{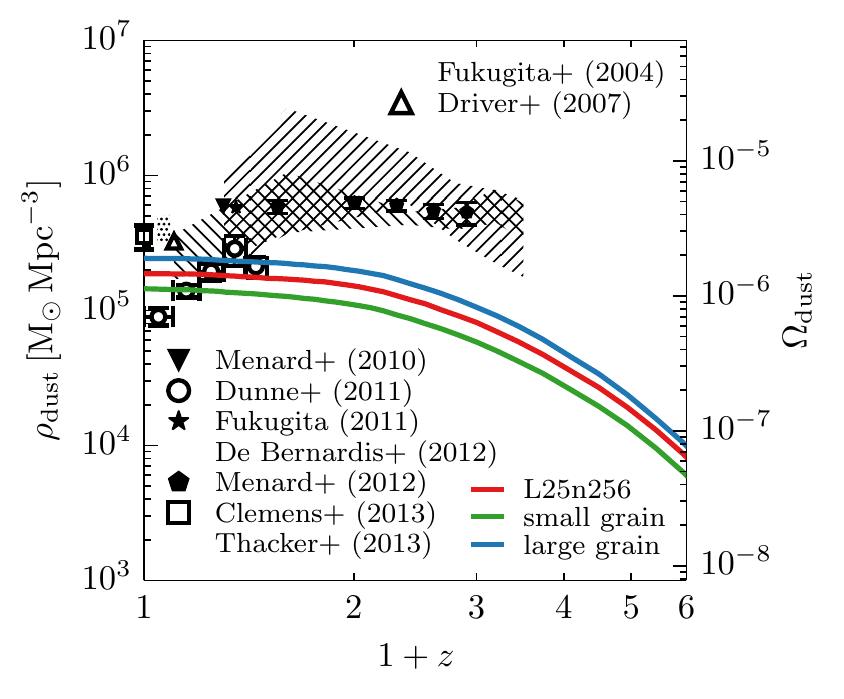}
\caption{Same as Figure~\ref{FIG:integrated_dust_density}, except using the
medium-resolution simulation with three grain size parameters.  The fiducial
L25n256 run uses $a = 0.1 \, \mu\text{m}$, while the small and large grain runs
use $a = 0.01 \, \mu\text{m}$ and $a = 1 \, \mu\text{m}$, respectively.
Evolution in the cosmic dust density is not sensitive to the choice of grain
size parameter.}
\label{FIG:integrated_dust_density_grain}
\end{figure}

The fiducial L25n256 run assumed a grain size of $a = 0.1 \, \mu\text{m}$ to
estimate thermal sputtering rates in Equation~(\ref{EQN:tau_sp}).  To investigate
the sensitivity of our results to this choice, we performed two additional runs
at the same resolution level: one with a smaller grain size ($a = 0.01 \,
\mu\text{m}$) and another with a larger grain size ($a = 1 \, \mu\text{m}$).
Figure~\ref{FIG:integrated_dust_density_grain} shows the cosmic dust density
and dust density parameter for $0 < z < 5$, which was previously studied in
Figure~\ref{FIG:integrated_dust_density}.

First, the results yield the correct qualitative behaviour: the sputtering
time-scale estimated in Equation~(\ref{EQN:tau_sp}) is longer for larger grains,
and we see that by $z = 0$ the cosmic dust density is largest for the $a = 1 \,
\mu\text{m}$ run and smallest for the $a = 0.01 \, \mu\text{m}$ run.
However, the dust densities predicted by these three runs differ by
less than a factor of two.  Figure~\ref{FIG:integrated_dust_density}
demonstrated that change in cosmic dust density when improving resolution from
the L25n128 to L25n256 run was just as large as varying the grain size
parameter by a factor of $100$.  Also, the observational data shown for
comparison indicate that there are larger uncertainties when estimating the
cosmic dust density through a variety of means (e.g.~quasar-galaxy reddening
correlations, power spectrum measurements, DMF integration, etc.).

Thus, the results presented in Section~\ref{SEC:results} are not sensitive to
our choice of $a$, especially when considering the combined uncertainties in
dust condensation efficiencies, dust mass absorption coefficients, and the
amount of dust in galactic haloes.

\section{Variation of Simulation Volume}\label{SEC:appendix_dmf}

\begin{figure}
\centering
\includegraphics{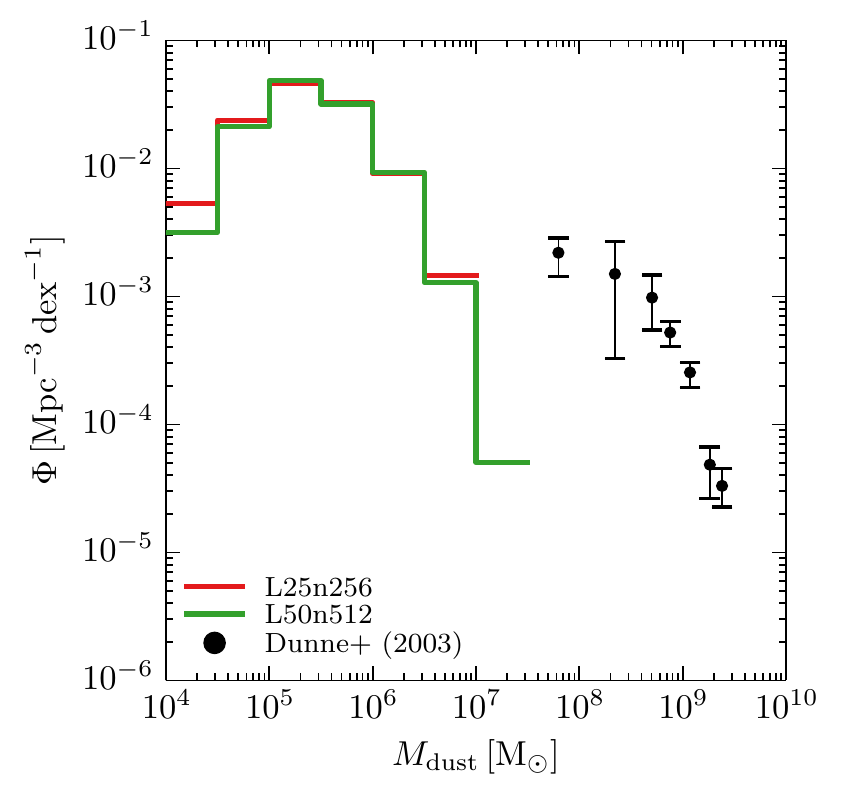}
\caption{Comparison of the $z = 2.5$ DMFs for the L25n256 run (red)
and the L50n512 run (green), the latter of which simulates a $(50 \,
h^{-1} \, \text{Mpc})^3$ volume with the same resolution as the L25n256 run.
This volume is eight times larger than the fiducial volume and enables us to
sample more galaxies.  The normalisation of the DMF is not substantially
changed by an increase in simulated volume.}
\label{FIG:dmf_bigbox}
\end{figure}

In addition to the fiducial runs, we also simulate a $(50 \, h^{-1} \,
\text{Mpc})^3$ volume down to $z = 2.5$ with $2 \times 512^3$ dark matter and
gas particles to start.  This run, labelled L50n512, uses the same fiducial
parameters from Table~\ref{TAB:parameters} and offers the same spatial and mass
resolution as the L25n256 run, but in a volume eight times as large.
Figure~\ref{FIG:dmf_bigbox} shows the DMFs for the L25n256 and L50n512 runs at
$z = 2.5$, which are nearly identical.  The number of galaxies in the
mass bin covering $10^{6.5} \, \text{M}_\odot \leq M_\text{dust} < 10^{7.0} \,
\text{M}_\odot$ has increased from $36$ in the L25n256 run to $254$ in the
L50n512 run, a change similar to the factor of eight increase in volume between
these runs.  The L50n512 run also forms $10$ galaxies in the next highest mass
bin, which had no galaxies in the L25n256 run.  Figure~\ref{FIG:dmf_bigbox}
suggests that the normalisation offset between the simulated DMFs and
observations is not the result of limited statistics.

\label{lastpage}

\end{document}